\documentclass[
 reprint,
 amssymb, amsmath,
 aip,jcp
]{revtex4-1}

\usepackage{comment}
\usepackage{dcolumn}
\usepackage[nolist]{acronym}

\usepackage[colorlinks=true,linkcolor=blue]{hyperref}%
\expandafter\ifx\csname package@font\endcsname\relax\else
 \expandafter\expandafter
 \expandafter\usepackage
 \expandafter\expandafter
 \expandafter{\csname package@font\endcsname}%
\fi
\usepackage{xspace}

\usepackage{graphicx}
\usepackage{amsmath}

\usepackage[normalem]{ulem}
\usepackage[utf8]{inputenc}

\usepackage{tikz}
\usepackage{multirow}
\usetikzlibrary{positioning}

\begin{acronym}
  \acro{FCI}{full configuration interaction}
  \acro{CI}{configuration interaction}
  \acro{QMC}{Quantum Monte Carlo}
  \acro{MO}{molecular orbital}
  \acro{HF}{Hartree-Fock}
  \acro{CAS}{complete active space}
  \acro{VMC}{variational Monte Carlo}
  \acro{DMC}{diffusion Monte Carlo}
  \acro{TC}{transcorrelated}
  \acro{FCIQMC}{full configuration interaction quantum Monte Carlo}
  \acro{SCF}{self consistent field}
  \acro{LM}{linear method}
  \acro{BLM}{blocked linear method}
  \acro{SR}{stochastic reconfiguration}
  \acro{GD}{gradient descent}
  \acro{ZV}{zero-variance}
\end{acronym}

\newcommand{\ket}[1]{{\ensuremath{|#1\rangle}\xspace}}
\newcommand{\bra}[1]{{\ensuremath{\langle #1|}\xspace}}
\newcommand{\braket}[2]{{\bra{#1} #2 \rangle\xspace}}
\newcommand{\elemm}[3]{\bra{#1} #2 \ket{#3}}
\newcommand{\basis}[0]{\mathcal{B}}

\newcommand{\phiiub}[0]{\ket{{\Phi_i^\basis}[u]}}
\newcommand{\chiiub}[0]{\ket{{\chi_i^\basis}[u]}}

\newcommand{\eib}[0]{\tilde{E}_i^\basis[u]}
\newcommand{\ei}[0]{\tilde{E}_i}

\newcommand{\phii}[0]{ \ket{{\Phi_i}}}

\newcommand{\pphii}[0]{{\Phi_i}}
\newcommand{\cchii}[0]{{\chi_i}}
\newcommand{\pphij}[0]{{\Phi_j}}
\newcommand{\cchij}[0]{{\chi_j}}

\newcommand{\br}[1]{{\mathbf{r}_{#1}}}

\newcommand{\brb}[1]{{\bf r}_{#1}}
\newcommand{\rab}[1]{|\brb{1} - \brb{2}|}

\newcommand{\ai}[1]{\hat{a}_{#1}}

\newcommand{\ku}[2]{\hat{K}[u](\br{#1},\br{#2})}
\newcommand{\kmu}[2]{\hat{K}[\mu](\br{#1},\br{#2})}
\newcommand{\lu}[3]{\hat{L}[u](\br{#1},\br{#2},\br{#3})}
\newcommand{\lmu}[3]{\hat{L}[\mu](\br{#1},\br{#2},\br{#3})}
\newcommand{\uu}[2]{u(\br{#1},\br{#2})}
\newcommand{\umu}[2]{u(r_{#1#2},\mu)}
\newcommand{\tu}{\hat{\tau}_u}
\newcommand{\tmu}{\hat{\tau}_\mu}
\newcommand{\deriv}[3]{\frac{\partial^{#3} #1}{\partial {#2}^{#3}}}

\newcommand{\wmuijkl}[4]{{w}_{ij}^{kl}}
\newcommand{\kijkl}[0]{{K}_{ij}^{kl}}
\newcommand{\lmuijmkln}[0]{{L}_{ijm}^{kln}}

\newcommand{\hmu}[0]{\tilde{H}[\mu]}
\newcommand{\htc}[0]{\tilde{H}}
\newcommand{\htcz}[0]{\tilde{H}_0}
\newcommand{\vtc}[0]{\tilde{V}}
\newcommand{\hu}[0]{\tilde{H}[u]}
\newcommand{\hub}[0]{\tilde{H}^\basis[u]}

\newcommand{\hmub}[0]{\tilde{H}^\basis[\mu]}
\newcommand{\adi}[1]{a^{\dagger}_{#1}}

\newcommand{\fderiv}[2]{\frac{\delta #1}{\delta{#2}}}
\newcommand{\ephi}[0]{E[\Psi]}
\newcommand{\etchiphi}[0]{\tilde{E}[\chi,\Phi]}
\newcommand{\etchiphis}[0]{\tilde{E}[\chi,\Phi^*]}
\newcommand{\etchi}[1]{\tilde{E}[\chi,#1]}
\newcommand{\etphi}[1]{\tilde{E}[#1,\Phi]}

\newcommand{\choiceae}[0]{E^{(0)}_{\text{coef}}}
\newcommand{\choicebe}[0]{E^{(0)}_{\text{sym}}}
\newcommand{\choicece}[0]{E^{(0)}_{\text{n sym}}}
\newcommand{\choiceh}[0]{E^{(0)}_{\text{regular H}}}
\newcommand{\choicehht}[0]{E^{(0)}_{\text{Hermit}}}
\newcommand{\choiceaept}[0]{(E^{(0)}+E^{(2)})_{\text{coef}}}
\newcommand{\choicebept}[0]{(E^{(0)}+E^{(2)})_{\text{sym}}}
\newcommand{\choicecept}[0]{(E^{(0)}+E^{(2)})_{\text{n sym}}}
\newcommand{\norm}[0]{\text{4-idx}}
\newcommand{\rmi}[0]{{\rm I}}
\newcommand{\rmj}[0]{{\rm J}}
\newcommand{\rmio}[0]{{\rm I}_0}

\usepackage{csquotes}

\begin{document}

\author{Abdallah Ammar}
\email{aammar@irsamc.ups-tlse.fr}
\newcommand{\LCPQ}{Laboratoire de Chimie et Physique Quantiques (UMR 5626), Universit\'e de Toulouse, CNRS, UPS, France}
\affiliation{\LCPQ}
\author{Anthony Scemama}
\affiliation{\LCPQ}
\author{Emmanuel Giner}%
\email{emmanuel.giner@lct.jussieu.fr}
\affiliation{Laboratoire de Chimie Théorique, Sorbonne Université and CNRS, F-75005 Paris, France}

\title{Extension of selected configuration interaction for transcorrelated methods}

\newcommand{\todo}[1]{{\color{red}TODO:#1}}

\begin{abstract}
In this work we present an extension of the popular selected configuration interaction (SCI) algorithms to the Transcorrelated (TC) framework.
Although we used in this work the recently introduced one-parameter correlation factor \href{https://doi.org/10.1063/5.0044683}{[E. Giner, J. Chem. Phys., 154, 084119 (2021)]}, the theory presented here is valid for any correlation factor.
Thanks to the formalization of the non Hermitian TC eigenvalue problem as a search of stationary points for a specific functional depending both
on left- and right-functions, we obtain a general framework allowing different choices for both the selection criterion in SCI and the second order perturbative correction to the energy.
After numerical investigations on different second-row atomic and molecular systems in increasingly large basis sets,
we found that taking into account the non Hermitian character
of the TC Hamiltonian in the selection criterion is mandatory to obtain a fast convergence of the TC energy.
Also, selection criteria based on either the first order coefficient or the second order energy lead to significantly
different convergence rates, which is typically not the case in the usual Hermitian SCI.
Regarding the convergence of the total second order perturbation energy, we find that the quality of the left-function used in the equations
strongly affects the quality of the results.
Within the near-optimal algorithm proposed here we find that the SCI expansion in the TC framework converges faster than the usual
SCI both in terms of basis set and number of Slater determinants.


\end{abstract}

\maketitle

\section{Introduction}
Obtaining an accurate description of the electronic structure of atomic and molecular systems
is the cornerstone of wave function theory (WFT) which aims at solving the many-body Schrödinger equation (SE)
for atomic and molecular systems.
An interesting feature of WFT is that the path toward the exact solution of the SE is \textit{a priori} known:
one has to compute the full configuration interaction (FCI) in increasingly large one-electron basis sets
until the complete basis set (CBS) limit is reached.
There are nevertheless two major drawbacks in WFT: i) the computational cost of FCI scales exponentially with the
number of electrons and number of basis functions, and ii) the convergence of most of the chemically relevant
properties, such as atomization energies or electrical responses, is slow with the size of the basis set.
The mixing of these two issues is such that the FCI is only applicable to few electron systems in moderate basis sets.
Therefore, intense efforts have been carried by the quantum chemistry community to develop new schemes
in order to alleviate these two problems.

The field of the so-called post Hartree Fock (HF) methods aims at developing efficient approximations to the FCI
wave function and energy within a given basis set starting from the HF mean-field solution.
There are many different ways of tackling the many-body problem which can be essentially split into two categories:
the variational approaches consisting in CI approaches, and the projective techniques relying
on perturbation theory (PT).
One advantage of variational approaches is that they are conceptually simple:
one explicitly builds the wave function as a linear combination of Slater determinants, whose coefficients and orbitals can
be optimized by minimization of the expectation value of the Hamiltonian.
The variational principle guarantees an upper bound to the exact ground state energy
which allows to treat strongly correlated systems without any divergence caused by near degeneracies and/or strong off
diagonal Hamiltonian matrix elements.
The major drawback of this approach is that the linear parametrization of the wave function
prohibits desirable properties such as size extensivity unless a complete active space (CAS) is chosen,
with a size scaling exponentially with the number of correlated electrons and orbitals.
On the other hand, the projective methods use two different wave functions to evaluate the energy without the burden of the usual expectation value:
a reference wave function which is typically a qualitative representation of the wave function (such as the HF Slater determinant),
and a correlated wave function which is a closer approximation to the FCI.
Thanks to the two-body nature of the Hamiltonian, the energy can be obtained
by essentially knowing the coefficients on the single- and double-particle-hole excitations
with respect to the reference wave function. Thanks to this simplification, PT can be developed
to produce useful computational tools such as M{\o}ller Plesset at second order\cite{mp} (MP2)
and important theorems\cite{brueckner,goldstone,lindgren} can be derived which allow the understanding of the product structure of the wave function.
The latter has led to the coupled cluster\cite{review_cc_bartlett} (CC)
exponential ansatz for the wave function guaranteeing size extensive energies and which can be obtained routinely at polynomial costs
thanks to intense developments on both the conceptual and practical aspects.
Despite the tremendous successes of approximate CC approaches, the drawback of such schemes are certainly their
difficulties to treat the strong correlation regimes where odd behaviours often occur because of near degeneracies
between several Slater determinants.
An alternative path has been proposed with the so-called selected-CI
\cite{bender,malrieu,buenker1,buenker-book,three_class_CIPSI,harrison,GinSceCaf-CJC-13,GinSceCaf-JCP-15,hbci,SchEva-JCTC-17,GinTewGarAla-JCTC-18,LooSceBloGarCafJac-JCTC-18,LooBogSceCafJac-JCTC-19,QP2,ZhaLieuHof-JCTC-21} (SCI) approaches
which can be thought as a mixing of the variational and projective methods.
While usual CI techniques predetermine the set of Slater determinants in the wave function, the SCI approaches aim at iteratively
selecting \emph{on-the-fly} the most relevant Slater determinants thanks to an importance criterion based on PT,
as initially proposed in the CI perturbatively selected iteratively (CIPSI)\cite{malrieu} of Malrieu and co-workers.
Thanks to this selection of the Slater determinants, the variational energy of the reference wave function converges rapidly towards the FCI energy.
Although the seminal works on SCI have been carried essentially between the seventies and nineties,
there exists a quite recent literature on SCI.
Nevertheless, most of these algorithms differ by the importance criterion used to select Slater determinants
(either the first order perturbed coefficient or second order contribution on the energy),
which leads to very similar convergence rates for the variational energy.
Another class of SCI methods applies a screening on the two-electron integrals in order to discard negligible excitations,
such as in EXSCI\cite{Bories_2007} or in the heath-bath-CI\cite{hbci} (HCI).
On-top of the variational energy of the reference wave function, one can add a second order perturbative correction on the energy, which allows
to drastically improve the convergence of the SCI algorithm towards the FCI energy.
Usually, the PT is carried with an Epstein-Nesbet\cite{epstein-PR-26,nesbet-PRCLA-55} (EN) zeroth order Hamiltonian in a multi-reference (MR) framework,
although attempts have been proposed to mix with the MP zeroth order Hamiltonian\cite{AngCimMal-CPL-00}.
Due to the important computational cost of the multi-reference perturbation theory (MRPT), different flavours of stochastic and semi-stochastic versions
of the latter were independently proposed in order to significantly speedup the calculations\cite{ShaHolJeaAlaUmr-JCTC-17,GarSceLooCaf-JCP-17}.
This recent renewal of the SCI techniques have pushed in a significant way the boundaries of accessible near FCI energies,
but they can give accurate estimates of the FCI energy only for systems with a few tens of correlated electrons in typically two hundreds of orbitals
\cite{benzene_bench,LooDamSce-JCP-20}.
The reason for such a limitation is the so-called exponential wall: even if the linear parametrization is compacted
thanks to efficient selection criteria
and further enhanced with a second order perturbative correction, it cannot compete with the exponentially growing number of determinants
in the FCI space.
Recent alternatives have been proposed to cure this problem by an exponential ansatz
using single reference CC with a selection \emph{a la CIPSI} of the individual excitation operators\cite{XuUejTen-JPCL-20,GurDeuShePie-JCP-21}.

Except for methods to approximate the FCI wave function in a given basis set, alternative tools have been proposed in order to
temper the slow convergence of the results of WFT with respect to the one-electron basis set.
The latter was acknowledged in the early years of quantum physics by Hylleraas\cite{Hyl-ZP-29}
to originate from the divergence of the Coulomb potential at short inter electronic distances,
which, as shown by Kato\cite{Kat-CPAM-57}, induces a derivative discontinuity (the so-called electron-electron cusp)
in the exact electron wave function.
Based on the latter results, there are essentially two branches that have emerged to alleviate the basis set convergence problems of WFT:
pure WFT approaches where one includes explicitly the inter electronic distances in the wave function,
and hybrid theories based on WFT and range-separated density functional theory\cite{TouColSav-PRA-04} (RSDFT).
The latter strategy, initially proposed in Ref.~\onlinecite{GinPraFerAssSavTou-JCP-18}, exploits the fact that the Coulomb potential
projected onto an incomplete basis set is non divergent, allowing for a mapping with RSDFT through the so-called long-range interaction used in this framework.
This hybrid scheme was successfully validated for the calculations of different chemically relevant properties including
light and transition metal atoms\cite{GinPraFerAssSavTou-JCP-18,LooPraSceTouGin-JCPL-19,GinSceTouLoo-JCP-19,GinSceLooTou-JCP-20,YaoGinTouUmr-JCP-20,LooPraSceGinTou-JCTC-20,YaoGinTylTouUmr-JCP-21,GinTraPraTou-JCP-21}.

Regarding pure WFT schemes dealing with the electron-electron cusp, the main idea is to introduce a so-called correlation factor which explicitly introduces
the inter electronic distances and there are mainly three branches of such methods which differ by the treatment of the correlation factor.
A first approach is variational Monte Carlo\cite{FouMitNeeRaj-RMP-01,AusZubLes-CR-12,TouAssUmr-INC-16} (VMC), 
where the wave function is expressed as the product of a Slater determinant expansion with a correlation factor. All the parameters of the wave function are optimized in a non-orthonormal stochastic framework.
The two main advantages of VMC are that virtually any form of correlation factor can be used as the $3N$-dimensional integrals
are computed with a Monte Carlo (MC) sampling, and also that the determinant expansion is strongly compacted by the correlation factor thanks to its non-negligible overlap with the Slater determinant basis.
Nonetheless, a major drawback of VMC originates from the statistical fluctuations of the sampled quantities needed to optimize
the wave function.

In the explicitly correlated methods\cite{rev_f12_tew,rev_f12_vallev,rev_f12_gruneis} (F12), the effect of the
correlation factor is projected outside the $N$-electron Hilbert space spanned by the finite one-electron basis set,
but one nevertheless obtains a faster convergence of the energy towards CBS results\cite{TewKlopNeiHat-PCCP-07}.
The F12 machinery induces numerous three- and four-electron integrals\cite{BarLoo-JCP-17} which constrains the correlation factor to have a rather simple form.
In addition, the presence of the correlation factor does not make the wave function expansion more compact, because the correlation factor is projected onto the Hilbert space orthogonal to that spanned by the basis set.

Another related approach consists in the so-called transcorrelated\cite{Hirschfelder-JCP-63,BoyHan-PRSLA-69,BoyHan-2-PRSLA-69} (TC) methods
where the correlation factor is introduced in the Hamiltonian instead of being introduced in the wave function.
The TC methodology proposed by Boys and Handy\cite{BoyHan-PRSLA-69,BoyHan-2-PRSLA-69} relies on a similarity transformation of the usual Hamiltonian by the correlation factor,
which necessarily leads to a non Hermitian operator but also maintains the orthonormality of the Slater determinant basis. 
It can be thought as a compromise between the F12 and VMC approaches:
as the full effect of the correlation factor is retained in the TC approach, the cuspless wave function expansion is compacted\cite{DobLuoAla-PRB-19,Gin-JCP-21,Sokolov2022},
and no more than three-electron integrals are needed in the optimization process.
After the seminal works by Boys and Handy\cite{BoyHan-PRSLA-69,BoyHan-2-PRSLA-69} where both the orbitals of a single Slater determinant and a rater sophisticated
correlation factor were optimized together, Ten-No\cite{TenNo-CPL-00-a} proposed a significant change of paradigm:
using a rather simple universal correlation factor shaped for valence electrons and giving more flexibility to the Slater determinant expansion to adapt to the presence
of the correlation factor.
This strategy was initially applied to MP2 in Refs.~\onlinecite{TenNo-CPL-00-a,HinTanTen-JCP-01} and to a linearized coupled cluster ansatz in later work\cite{HinTanTen-CPL-02}.
Attempts to remain within a variational framework despite the non Hermitian nature of the TC Hamiltonian have been proposed by
Umezawa \textit{et. al.}\cite{UmeTsu-JCP-03,UmeTsuOhnShiChi-JCP-05} and Luo\cite{Luo-JCP-10,Luo-JCP-11}, and an alternative similarity transformation with an anti Hermitian
correlation factor was proposed by Yanai \textit{et. al.}\cite{NeuYanCha-MolPhys-10,YanShi-JCP-12}.
Developments of the TC method towards the treatment of solid state systems have been carried by Ochi
\textit{et. al.} including both ground\cite{OchSodSakTsu-JCP-12,OchTsu-JCTC-14,OchTsu-CPL-15,OchYamAriTsu-JCP-16} and
excited states\cite{OchTsu-JCTC-14}.
More recently, Cohen \textit{et. al}\cite{CohLuoGutDobTewAla-JCP-19,GutCohLuoAla-JCP-21} applied the TC equations with an elaborate correlation factor
and proposed to use the full configuration interaction Monte Carlo (FCIQMC) method to obtain the exact ground state energy and the corresponding right eigenvector of the TC Hamiltonian in a given basis set.
Their approach, the so-called TC-FCIQMC, originally used the correlation factors published by Moskowitz \textit{et. al.}\cite{SchMos-JCP-90}
which were optimized in the context of VMC for the He-Ne neutral series, and which explicitly take into account electron-electron-nucleus correlation effects.
Adaptations of the CC equations to the TC framework were also applied to molecular\cite{SchCohAla-JCP-21} and periodic systems\cite{KeSchLuoKatAla-PRR-21}.
Applications of the similarity transformation for Hubbard model Hamiltonians with the
Gutzwiller Ansatz for the correlation factor\cite{Gutzwiller-PRL-63,BriRic-PRB-70}
were carried using FCIQMC\cite{DobLuoAla-PRB-19} and density matrix renormalization group\cite{BaiRei-JCP-20}.
Adaptations of the TC equations to matrix product state methodology have been also reported\cite{SchCohAla-JCP-21,BaiLesRei-JCTC-22}.

Recently, one of the present authors\cite{Gin-JCP-21} introduced a correlation factor which generates, at leading order in $1/r_{12}$,
an effective potential in the TC Hamiltonian reproducing the non divergent long range interaction of RSDFT.
The correlation factor only depends on the inter electronic distance, and is tuned by a single parameter $\mu$ which controls
both the range and depth of the correlation hole induced in the cuspless wave function.
The total effective potential appearing in the TC Hamiltonian has a relative simple analytical structure and the two- and three-electron
integrals can be obtained efficiently using a mixed numerical/analytical scheme.
The first applications on two electron systems in Ref.~\onlinecite{Gin-JCP-21} have shown encouraging results on the convergence of the energy
and the compaction of the wave function, and also enabled a rather systematic way to obtain a reasonable system-dependent
value for the parameter $\mu$ depending only on the HF density.
Coupled with the TC-FCIQMC methodology, further tests on the Li-Ne neutral and first-cation series together with second-row molecular systems\cite{DobCohAlaGin-JCP-22}
have shown that such a correlation factor in the context of TC equations is competitive with the rather complex correlation
factors used by Alavi \textit{et. al.} in Ref.~\onlinecite{CohLuoGutDobTewAla-JCP-19}.

The present work proposes to adapt the SCI strategy to the general TC framework in order to benefit from the compaction of the wave function
due to the presence of the correlation factor and to speed up the convergence of the results with respect to the basis set.
We use here the $\mu$-dependent correlation factor of Ref.~\onlinecite{Gin-JCP-21}, but the equations derived here are general and applicable to
any correlation factor. Therefore the main goal of the present approach is more the specificity of the SCI strategy in the TC context rather than
the actual performance of the $\mu$-dependent correlation factor.

The paper is organized as follows. In Sec.~\ref{sec:theo_tc} we briefly recall the form of the general TC equations in first- and second-quantization.
Different levels of approximations for the three-body terms used here are presented in Sec.~\ref{sec:three_b} and the TC equations
with the $\mu$-dependent correlation factor used here are briefly exposed in Sec.~\ref{sec:mu_tc}.
Then, in Sec.~\ref{sec:bilinear} we expose the connection between non Hermitian eigenvalue problems
and the stationary points of a functional depending on two functions,
which allows us then to expose a perturbative expansion of such a formulation in Sec.~\ref{sec:pt_bilin}.
Based on these tools, we extend the SCI problem to non Hermitian Hamiltonians in Sec.~\ref{sec:nh_sci},
and also present in Sec.~\ref{sec:diag_nh} a framework to solve non Hermitian eigenvalue problems with only Hermitian matrices.
In Sec.~\ref{sec:results} we present the numerical results obtained.
We investigate on several atomic and one molecular systems the convergence of the different flavours of SCI
in the TC context in Sec.~\ref{sec:conv_sci} in order to determine the most efficient way of performing SCI.
Having established a near optimal strategy, in Sec.~\ref{sec:res_three_body} we investigate the dependency
of both total energies and atomization energies on the level of treatments of the three-body terms on a set of atomic and molecular systems.
Eventually, we summarize and conclude in Sec.~\ref{sec:conclu}.

\section{Theory}
\label{sec:theo}
\subsection{General equations and concepts of TC theory}
\label{sec:theo_tc}
The general form of the transcorrelated Hamiltonian for a symmetric correlation factor $\uu{1}{2}$ is given by
\begin{equation}
 \label{ht_def_g}
 \begin{aligned}
  \hu  &\equiv e^{-\tu} \hat{H} e^{\tu} \\
                & = H + \big[ H,\tu \big] + \frac{1}{2}\bigg[ \big[H,\tu\big],\tu\bigg],
 \end{aligned}
\end{equation}
where $\tu = \sum_{i<j}u(\br{i},\br{j})$ and $\hat{H} = -\sum_i \frac{1}{2} \nabla^2_i + v(\br{}_i) + \sum_{i<j}   1/r_{ij}$.
Eq. \eqref{ht_def_g} leads to the following transcorrelated Hamiltonian
\begin{equation}
 \begin{aligned}
 \label{ht_def_g2}
 \hu& = H - \sum_{i<j} \ku{i}{j} - \sum_{i<j<k} \lu{i}{j}{k},
 \end{aligned}
\end{equation}
where the effective two- and three-body operators $\ku{1}{2}$ and $\lu{1}{2}{3}$ are defined as
\begin{equation}
 \begin{aligned}
  \ku{1}{2} = \frac{1}{2} \bigg( &\Delta_1 \uu{1}{2} + \Delta_2 \uu{1}{2} \\
                                         + &\big(\nabla_1 \uu{1}{2} \big) ^2 + \big(\nabla_2 \uu{1}{2}      \big) ^2 \bigg) \\
                                         + &\nabla_1 \uu{1}{2} \cdot \nabla_1 + \nabla_2 \uu{1}{2}\cdot     \nabla_2,
 \end{aligned}
\end{equation}
and
\begin{equation}
 \begin{aligned}
  \lu{1}{2}{3} = &   \nabla_1 \uu{1}{2} \cdot \nabla_1 \uu{1}{3} \\
                                          + & \nabla_2 \uu{2}{1} \cdot \nabla_2 \uu{2}{3}   \\
                                          + & \nabla_3 \uu{3}{1} \cdot \nabla_3 \uu{3}{2}   .
 \end{aligned}
\end{equation}
In practice, the TC Hamiltonian is projected into a one-particle basis set $\basis$
\begin{equation}
 \label{eq:def_hub}
 \begin{aligned}
 &\hub   = P^{\basis} \hu P^\basis,
 \end{aligned}
\end{equation}
where $P^{\basis}$ is the projector onto the $N_e$-electron Hilbert space spanned by the one-particle basis set $\basis$ 
and $N_e$ is the number of electrons.
Using real-valued orthonormal spatial molecular orbitals (MOs) $\{\phi_i(\br{})\}$, $\hub$ can be written in a second-quantized form  as
\begin{equation}
 \label{eq:def_hub_sec_q}
 \begin{aligned}
 &\hub  = \sum_{i,j \in \basis} \,\, \sum_{\sigma = \uparrow,\downarrow} h_{ij} \adi{j,\sigma}\ai{i,\sigma}\\
 & + \frac{1}{2}\sum_{i,j,k,l \in \basis}  \,\, \sum_{\sigma,\lambda = \uparrow,\downarrow}
 \big( V_{ij}^{kl} - \kijkl\big) \adi{k,\sigma} \adi{l,\lambda} \ai{j,\lambda} \ai{i,\sigma} \\
       & - \frac{1}{6} \sum_{i,j,m,k,l,n \in \basis} \,\, \sum_{\sigma,\lambda,\kappa = \uparrow,\downarrow}
\lmuijmkln \adi{k,\sigma} \adi{l,\lambda} \adi{n,\kappa} \ai{m,\kappa} \ai{j,\lambda} \ai{i,\sigma},
 \end{aligned}
\end{equation}
where $h_{ij}$ are the usual one-electron integrals, $V_{ij}^{kl}$ are the usual two-electron integrals,
$\kijkl$ are the two-electron integrals corresponding to the effective two-body operator $\ku{1}{2}$ 
\begin{equation}
 \kijkl = \int \text{d} \br{1} \text{d} \br{2} \phi_k(\br{1}) \phi_l(\br{2}) \ku{1}{2} \phi_i(\br{1}) \phi_j(\br{2}),
\end{equation}
and $\lmuijmkln$ are the three-electron integrals corresponding to the effective three-body operator $\lu{1}{2}{3}$
\begin{equation}
 \begin{aligned}
  \lmuijmkln  = \int \text{d} \br{1} \text{d} \br{2} \text{d} & \br{3} \phi_k(\br{1}) \phi_l(\br{2}) \phi_n(\br{3}) \\ & \lu{1}{2}{3} \phi_i(\br{1}) \phi_j(\br{2}) \phi_m(\br{3}).
 \end{aligned}
\end{equation}
Since $\hub$ is non Hermitian, a given eigenvalue $\eib$ can be associated with a couple of right- and left-eigenvectors
\begin{equation}
 \label{eq:tc_eigv}
 \begin{aligned}
 & \hub \phiiub = \eib \phiiub \\
 & \big(\hub\big)^\dagger \chiiub = \eib \chiiub,\\
 \end{aligned}
\end{equation}
and because of the properties of the similarity transformation the exact eigenvalue $E_i$ is recovered in the CBS limit
\begin{equation}
 \lim_{\basis \rightarrow \text{CBS}} \eib = E_i.
\end{equation}
As a part of the correlation effects are taken into account with the correlation factor, one can expect that the convergence of $\eib$ is more rapid than the usual WFT-based method.

Since all calculations presented here are performed in an incomplete basis set $\basis$, from hereon we omit the basis set $\basis$ symbol for the sake of the simplicity of the notations.

\subsection{Approximations for three body terms}
\label{sec:three_b}
The computation and storage of the 6-index $\lmuijmkln$ tensor corresponding to the three-body operator $\lu{1}{2}{3}$ rapidly becomes
the main computational bottleneck in the TC calculations.
To overcome such limitations, several approximations to the full treatment of the three-body terms have been proposed in the
literature\cite{HinTanTen-JCP-01,HinTanTen-CPL-02,SchCohAla-JCP-21,DobCohAlaGin-JCP-22}.
When using methods such as MP2 or CC approaches for which the needed integrals are known \textit{a priori} of the calculations,
one can use a resolution of the identity approximation (RI) as proposed in Ref.~\onlinecite{HinTanTen-JCP-01}, with only
$N^4$-storage requirement for intermediate quantities.
Nevertheless, the RI would be very costly to be used in the context of SCI or stochastic approaches such as FCIQMC
as one cannot anticipate which specific integrals will be involved in matrix elements and it would therefore
imply to recompute the three-electron integrals whenever they are required in the matrix elements computations.

In Refs.~\onlinecite{HinTanTen-CPL-02,SchCohAla-JCP-21} the authors proposed to use the normal-ordering of the three-body term on
a reference determinant in order to obtain effective  one- and two-body operators,
which results in a typical $N^4$ computational scaling but nevertheless introduces a dependence on the reference determinant chosen
for the normal ordering.
In Ref.~\onlinecite{DobCohAlaGin-JCP-22} the authors introduced the so-called ``5-idx'' approximation consisting in simply neglecting
all $\lmuijmkln$ integrals with six different indices, which results in a less-favourable $N^5$ scaling but does not introduce
an explicit dependence on the chosen reference determinant. Nevertheless, the three-body operator being truncated, it introduces necessarily a dependence on the MO basis used, but numerical investigations carried in  Ref.~\onlinecite{DobCohAlaGin-JCP-22} have shown that this dependency is rather small.
Here we propose to introduce a new approximation which can be seen as a compromise between the normal-ordering technique and the 5-idx approximation.
Such approximation, here referred to as $\norm$, consists in treating explicitly the three-body terms for diagonal
and single-excitation matrix elements (which therefore account for all $\lmuijmkln$ involving at most four different indices)
and using the two-body sector of the normal order operator with four different indices for the treatment of the double excitations Hamiltonian
matrix elements. The $\norm$ approximation results in a typical $N^4$ computational scaling but as the explicit treatment of the
three-body terms are retained for the diagonal and single-excitation matrix elements, it might mitigate the dependence on the reference determinant.
The 5-idx approximation is used throughout this work unless the $\norm$ is explicitly mentioned.

\subsection{One-parameter TC Hamiltonian: $\hmu$}
\label{sec:mu_tc}
The one-parameter correlation factor $\umu{1}{2}$ with $r_{12} = |\mathbf{r}_1 - \mathbf{r}_2|$ used in this work was originally proposed in Ref.~\onlinecite{Gin-JCP-21}
based on a mapping between the $r_{12} \approx 0$ limit of the TC Hamiltonian and the RSDFT effective Hamiltonian.
The explicit form of the correlation factor $\umu{1}{2}$  is given by
\begin{equation}
 \label{eq:def_j}
 \umu{1}{2} = \frac{1}{2}r_{12}\bigg( 1 - \text{erf}(\mu r_{12})  \bigg) - \frac{1}{2\sqrt{\pi}\mu}e^{-(r_{12}\mu)^2}.
\end{equation}
Because of the simple analytical expression of $\umu{1}{2}$, the corresponding TC Hamiltonian  $\hmu$ defined as
\begin{equation}
 \begin{aligned}
 \hmu &\equiv e^{-\tmu} \hat{H} e^{\tmu} \\
      & = H - \sum_{i<j} \kmu{i}{j} - \sum_{i<j<k} \lmu{i}{j}{k},
 \end{aligned}
\end{equation}
with $\tmu = \sum_{i<j}\umu{i}{j}$, has a relatively simple analytical form
with the effective two- and three-body operators
\begin{equation}
 \label{eq:k_final}
  \begin{aligned}
   & \kmu{i}{j}=  \frac{1 - \text{erf}(\mu r_{12})}{r_{12}} - \frac{\mu}{\sqrt{\pi}} e^{-\big(\mu       r_{12} \big)^2} \\
                &+ \frac{\bigg(1 -     \text{erf}(\mu r_{12}) \bigg)^2}{4} - \bigg( \text{erf}(\mu r_{12}) - 1\bigg) \deriv{}{r_{12}}{}
  \end{aligned}
\end{equation}
and
\begin{equation}
 \label{eq:l_final}
 \begin{aligned}
 \lmu{i}{j}{k} = & \frac{1 - \text{erf}(\mu r_{12})}{2 r_{12}} \br{12} \cdot \frac{1 -\text{erf}(\mu r_{13})}{2 r_{13}} \br{13} \\
               + & \frac{1 - \text{erf}(\mu r_{21})}{2 r_{21}} \br{21} \cdot \frac{1 -\text{erf}(\mu r_{23})}{2 r_{23}} \br{23} \\
               + & \frac{1 - \text{erf}(\mu r_{31})}{2 r_{31}} \br{31} \cdot \frac{1 -\text{erf}(\mu r_{32})}{2 r_{32}} \br{32},
 \end{aligned}
\end{equation}
respectively.
The correlation factor $\umu{1}{2}$ exactly restores the cusp conditions and the scalar two- and three-body effective interaction in Eqs.~\eqref{eq:k_final} and~\eqref{eq:l_final} lead  to a non divergent interaction in $\hmu$ which has then \enquote{cusp\-less} eigenvectors as illustrated in Ref~\onlinecite{Gin-JCP-21}.
As apparent from the definitions of Eq~\eqref{eq:k_final} and Eq.~\eqref{eq:l_final}, the global shape of $\hmu$ depends on a unique parameter $\mu$, which can be seen either as the inverse of the typical range of the correlation effects, or the typical value of the effective interaction at $r_{12}=0$.
In the $\mu \rightarrow +\infty$ limit one obtains the usual Hamiltonian, while in $\mu \rightarrow 0 $ limit one obtains a well defined attractive non Hermitian Hamiltonian even if the correlation factor becomes singular.

\subsection{Non Hermitian eigenvalue problems as stationary points of functional}
\label{sec:n_h_theo}
The present section describes how non Hermitian eigenvalue problems
can be rewritten in terms of stationary points of functionals depending on two functions.
We give the derivations for a general non Hermitian operator $\tilde{H}$ with real eigenvalues, \textit{i.e.} a pseudo-Hermitian operator as introduced in Ref.~\onlinecite{Mostafazadeh-JCP-02}.
It is worth noticing that according to the definition of
Mostafazadeh\cite{Mostafazadeh-JCP-02}, any operator $\hat{O}$ fulfilling
\begin{equation}
 \hat{O} = \hat{R}^{-1} \hat{O}^\dagger \hat{R},
\end{equation}
is a pseudo Hermitian operator (\textit{i.e.} with real eigenvalues), which is
of course the case of the TC Hamiltonian as
\begin{equation}
 (\hu)^\dagger = e^{+\tu} \hat{H} e^{-\tu},
\end{equation}
and therefore
\begin{equation}
 \hu = \hat{R}^{-1} (\hu)^\dagger  \hat{R},
\end{equation}
with $\hat{R} = e^{2\tu}$.
\subsubsection{Functionals and non Hermitian eigenvalue problems}
\label{sec:bilinear}
We consider here a non Hermitian operator $\tilde{H}$ with real eigenvalues.
Its left- and right-eigenvectors are real and not orthonormal
\begin{equation}
 \begin{aligned}
 & \braket{\pphii}{\pphij} \equiv S^r_{ij} \ne \delta_{ij}, \\
 & \braket{\cchii}{\cchij} \equiv S^l_{ij} \ne \delta_{ij}, \\
 \end{aligned}
\end{equation}
but can be rescaled such that they verify a bi-orthonormality relation
\begin{equation}
 \begin{aligned}
 & \braket{\cchii}{\pphij} = \delta_{ij}.
 \end{aligned}
\end{equation}
As a consequence, the usual energy functional
\begin{equation}
 \label{e_phi}
 E[\Psi] = \frac{\elemm{\Psi}{\htc}{\Psi}}{\braket{\Psi}{\Psi}},
\end{equation}
does not necessarily admit a lower bound. Indeed, assuming that $\braket{\Psi}{\Psi}=1$, one can expand the function $\ket{\Psi}$ on the right-eigenvectors for instance
\begin{equation}
 \ket{\Psi} = \sum_i c_i \phii,\quad c_i \in \mathbb{R},
\end{equation}
then $E[\Psi]$ can be expressed as
\begin{equation}
 \ephi = \sum_{i,j} c_i c_j E_i S^r_{ij},
\end{equation}
which has no reason to be an upper bound to $E_0$ as $S^r_{ij} \ne \delta_{ij}$.
Therefore looking for a minimum of the functional $E[\Phi]$ is irrelevant due to the loss of the variational principle.

Computing the functional derivative of Eq. \eqref{e_phi}
\begin{equation}
 \fderiv{\ephi}{\Psi} =
 \frac{ \braket{\Psi}{\Psi} \big( \htc + \htc^\dagger \big) \ket{\Psi} - 2 \elemm{\Psi}{\htc}{\Psi}\ket{\Psi}}{|\braket{\Psi}{\Psi}|^2},
\end{equation}
and looking for a stationary point $\ket{\Psi^*}$ normalized to unity such that
\begin{equation}
 \fderiv{E[\Psi^*]}{\Psi^*} = 0,
\end{equation}
yields the following eigenvalue equation
\begin{equation}
 \label{eq:eigv_sym}
 \frac{1}{2}\big( \htc + \htc^\dagger \big) \ket{\Psi^*} = \lambda \ket{\Psi^*}.
\end{equation}
Therefore looking for stationary points of $\ephi$ leads to eigenvectors of the symmetrized operator $\htc + \htc^\dagger$,
which are of course different from the eigenvectors of the original non Hermitian operator $\htc$.

In order to obtain a functional whose stationary points are the eigenvectors of $\htc$ one needs to define the
following functional
\begin{equation}
 \label{e_phichi}
 \etchiphi = \frac{\elemm{\chi}{\htc}{\Phi}}{\braket{\chi}{\Phi}},
\end{equation}
where the functions $\chi$ and $\Phi$ are called here the left- and right-function, respectively.
Being a functional of two functions, $\etchiphi$ admits two functional derivatives
\begin{equation}
 \label{deriv_chi}
 \fderiv{\etchiphi}{\chi} = \frac{\htc \ket{\Phi}\braket{\chi}{\Phi} - \elemm{\chi}{\htc}{\Phi} \ket{\Phi}}{|\braket{\chi}{\Phi}|^2},
\end{equation}
\begin{equation}
 \label{deriv_phi}
 \fderiv{\etchiphi}{\Phi} = \frac{\htc^\dagger \ket{\chi}\braket{\chi}{\Phi} - \elemm{\chi}{\htc}{\Phi} \ket{\chi}}{|\braket{\chi}{\Phi}|^2}.
\end{equation}
Therefore, if one searches for a stationary point $\ket{\Phi^*}$ such that, for a given $\ket{\chi}$,
the functional derivative vanishes
\begin{equation}
 \label{deriv_chi0}
 \fderiv{\etchiphis}{\chi} = 0,
\end{equation}
one obtains
\begin{equation}
 \label{deriv_chi0_phi}
 \htc \ket{\Phi^*} = \etchi{\Phi^*} \ket{\Phi^*},
\end{equation}
and one immediately recognizes the eigenvalue equations of the TC Hamiltonian for the right-eigenvector (see Eq. \eqref{eq:tc_eigv}).
Therefore, cancelling the left functional derivative yields an equation for the right-eigenvector $\ket{\pphii}$ of $\htc$,
which can seem counter intuitive.
This can be understood by noticing that when the functional $\etchiphi$ is evaluated at a right-eigenvector $\ket{\pphii}$,
the functional $\etchi{\pphii}$ is then insensitive to the function $\chi$
\begin{equation}
 \etchi{\pphii} = \frac{\elemm{\chi}{\htc}{\pphii}}{\braket{\chi}{\pphii}} = \ei\,\, \forall \chi,
\end{equation}
which is the definition of a stationary point with respect to the left function $\chi$.

Of course, similar equations can be obtained for the left-eigenvector
\begin{equation}
 \label{deriv_phi0}
 \fderiv{\etchiphi}{\Phi} = 0 \Leftrightarrow \htc^\dagger \ket{\chi} = \etphi{\chi^*}  \ket{\chi},
\end{equation}
and in the case where $\ket{\chi} = \ket{\chi_i}$ one obtains that
\begin{equation}
 \etphi{\cchii} = \ei \,\, \forall \Phi,
\end{equation}
which means that the value of the functional $\etphi{\cchii}$ becomes insensitive to $\Phi$ when evaluated at $\ket{\chi} = \ket{\chi_i}$.

\subsubsection{Perturbation theory of the functional}
\label{sec:pt_bilin}
Having in mind that finding right- (left-)eigenvectors is equivalent to find a stationary point of the functional $\etchiphi$,
one can then Taylor expand such a functional in order to obtain a perturbative expansion.
For the sake of simplicity of notations, we will omit the index labelling the state and implicitly focus on the ground state.

Let the Hamiltonian $\htc$ be split into
\begin{equation}
 \htc = \htcz + \lambda \vtc,
\end{equation}
and we would like to obtain the ground state energy $\tilde{E}_0$ as a Taylor expansion of the functional
$\etchi{\Phi_0}$ evaluated at a given $\ket{\chi}$ with
\begin{equation}
 \label{eq:pert_e_0}
  \etchi{\Phi_0}=  \frac{\elemm{\chi}{\htcz + \lambda \vtc }{\Phi_0}}{\braket{\chi}{\Phi_0}} 
                =  \sum_{k=0}^\infty \lambda^k \tilde{E}^{(k)},
\end{equation}
which therefore implies to also Taylor expand the right-eigenvector in powers of $\lambda$
\begin{equation}
 \ket{\Phi_0} = \sum_{k=0}^\infty \lambda^k \ket{\Phi^{(k)}},
\end{equation}
which satisfies the eigenvalue equation
\begin{equation}
 \label{eq:pert_phi_0}
 (\htcz + \lambda \vtc) \ket{\Phi_0} = \etchi{\Phi_0} \ket{\Phi_0},
\end{equation}
and we will start the expansion of $\ket{\Phi_0}$ from an eigenvector of $\htcz$, called here $\ket{\Phi^{(0)}}$
\begin{equation}
 \begin{aligned}
 \htcz \ket{\Phi^{(0)}} = \tilde{\epsilon}_0 \ket{\Phi^{(0)}}.
 \end{aligned}
\end{equation}
We will assume that the ground state $\ket{\Phi_0}$ can be expanded on a set of $N$ \textit{orthonormal} functions $\mathcal{S} = \{ \ket{\phi_i},\,i=1,N \}$
\begin{equation}
 \begin{aligned}
 &  \ket{\Phi_0} = \ket{\Phi^{(0)}} + \sum_{i} c_i \ket{\phi_i}, \\
 & \braket{\phi_j}{\phi_i} = \delta_{ij}, \\
 \end{aligned}
\end{equation}
which are also orthogonal to $\ket{\Phi_0}$
\begin{equation}
 \label{eq:ortho_p0_s}
  \braket{\Phi^{(0)}}{\phi_i} =  0\,\, \forall \ket{\phi_i}\in \mathcal{S},
\end{equation}
and for which $\htcz$ is \textit{diagonal} on $\mathcal{S}$
\begin{equation}
 \label{eq:def_h_0}
 \htcz \ket{\phi_i} = \epsilon_i^{(0)} \ket{\phi_i},
\end{equation}
implying that
\begin{equation}
 \begin{aligned}
  &\elemm{\Phi^{(0)}}{\htcz}{{\phi_i}} = \elemm{{\phi_i}}{\htcz}{\Phi^{(0)}} = 0 \,\, \forall \ket{\phi_i} \in \mathcal{S}.
 \end{aligned}
\end{equation}
Because we assumed that the functions $\ket{\phi_i}$ are orthonormal, the condition Eq. \eqref{eq:def_h_0} necessarily implies that $\htcz$ is \textit{Hermitian} on the set $\mathcal{S}$.
We will further assume that the set of $\ket{\phi_i}$ is orthogonal to the vector $\ket{\chi}$
\begin{equation}
 \braket{\chi}{\phi_i} = 0\,\,\forall \ket{\phi_i} \in \mathcal{S},
\end{equation}
which implies that
\begin{equation}
 \braket{\chi}{\Phi_0} = \braket{\chi}{\Phi^{(0)}}.
\end{equation}
Compared to the general case, this restriction considerably simplifies the perturbative expansion, and
will actually be relevant in the SCI framework we are interested in.
The orthonormality condition of Eq. \eqref{eq:ortho_p0_s} also implies that
\begin{equation}
 \braket{\Phi^{(0)}}{\Phi_0} = 1.
\end{equation}

Once the properties of $\htcz$ are defined on the whole space, one can replace
the expression of $\ket{\Phi_0}$ in Eq. \eqref{eq:pert_e_0} to obtain the perturbative expansion of the energy.
For the present purpose we stop at second order
\begin{equation}
 \label{eq:def_e_right_1}
 \begin{aligned}
  \etchi{\Phi_0}= &  \tilde{E}^{(0)} + \lambda  \tilde{E}^{(1)} +  \lambda^2 \tilde{E}^{(2)},  \\
                = & \frac{\elemm{\chi}{\htcz}{\Phi^{(0)}} + \lambda
 \elemm{\chi}{\vtc}{\Phi^{(0)}} + \lambda^2 \elemm{\chi}{\vtc}{\Phi^{(1)}}  }{\braket{\chi}{\Phi^{(0)}}} \\
 \end{aligned}
\end{equation}
which obviously gives
\begin{equation}
 \label{eq:def_e_right_2}
 \begin{aligned}
 &  \tilde{E}^{(0)} = \frac{\elemm{\chi}{\htcz}{\Phi^{(0)}}}{\braket{\chi}{\Phi^{(0)}}}, \\
 &  \tilde{E}^{(1)} = \frac{\elemm{\chi}{\vtc}{\Phi^{(0)}}}{\braket{\chi}{\Phi^{(0)}}}, \\
 &  \tilde{E}^{(2)} = \frac{\elemm{\chi}{\vtc}{\Phi^{(1)}}}{\braket{\chi}{\Phi^{(0)}}}.
 \end{aligned}
\end{equation}
To obtain the equation for the perturbed wave function one replaces the expressions of both
$\ket{\Phi_0}$ and $\etchi{\Phi_0}$ in Eq. \eqref{eq:pert_phi_0}
\begin{equation}
 \left(\htcz + \lambda \vtc \right) \left(\sum_{k=0}^\infty \lambda^k\ket{\Phi^{(k)}} \right) =
 \left( \sum_{k=0}^\infty \lambda^k \tilde{E}^{(k)}\right) \left(\sum_{m=0}^\infty \lambda^m\ket{\Phi^{(m)}} \right),
\end{equation}
and for $\Phi^{(1)}$ one obtains
\begin{equation}
 \label{eq:pert_phi1}
 \htcz \ket{\Phi^{(1)}} + \vtc \ket{\Phi^{(0)}} = \tilde{E}^{(0)} \ket{\Phi^{(1)}} + \tilde{E}^{(1)} \ket{\Phi^{(0)}}.
\end{equation}
By projecting Eq. \eqref{eq:pert_phi1} on a function $\ket{\phi_i}$ one obtains
\begin{equation}
 \label{eq:phi_1}
 c_i^{(1)} = \frac{\elemm{\phi_i}{\vtc}{\Phi^{(0)}}}{\tilde{E}^{(0)} - \epsilon_i},
\end{equation}
and therefore one can obtain the second order contribution to the energy
\begin{equation}
 \tilde{E}^{(2)} = \sum_{i=1}^N \tilde{E}^{(2)}_i,
\end{equation}
where $\tilde{E}^{(2)}_i $ is the contribution at second order to the energy of the function $\ket{\phi_i}$
\begin{equation}
 \label{eq:def_pt2}
 \begin{aligned}
 \tilde{E}^{(2)}_i & = \frac{\elemm{\chi}{\vtc}{\phi_i}\,c_i^{(1)}}{\braket{\chi}{\Phi^{(0)}}}  \\
                   & = \frac{1}{\braket{\chi}{\Phi^{(0)}}}\frac{\elemm{\chi}{\vtc}{\phi_i} \elemm{\phi_i}{\vtc}{\Phi^{(0)}} }{\tilde{E}^{(0)} - \epsilon_i}.
 \end{aligned}
\end{equation}
One should notice in Eq. \eqref{eq:def_pt2} that the non Hermitian nature of $\htc$ manifests in two ways: i) the coefficient is computed using $\elemm{\phi_i}{\vtc}{\Phi^{(0)}}$ which can be different from $\elemm{\Phi^{(0)}}{\vtc}{\phi_i}$,
and ii) the computation of the energy implies, in the general case, the use of another left-function $\chi$ through $\elemm{\chi}{\vtc}{\phi_i}$.

One can also Taylor expand the left-eigenvector $\ket{\chi_0}$ and obtain an alternative expression for the ground state eigenvalue
for a given function $\ket{\Phi}$
\begin{equation}
 \begin{aligned}
 &  \ket{\chi_0} = \sum_{k=0}^\infty \lambda^k \ket{\chi^{(k)}} \\
 &  \tilde{E}_{0}  = \sum_{k=0}^\infty \lambda^k \bar{E}^{(k)} \\
 &  \ket{\chi^{(k)}} = \sum_{i=1}^N \bar{c}_i^{(k)} \ket{\phi_i},\\
 \end{aligned}
\end{equation}
where we assume that $\ket{\chi^{(0)}}$ is a left eigenvector of $\htcz$
\begin{equation}
 \begin{aligned}
 \htcz^\dagger \ket{\chi^{(0)}} = \tilde{\epsilon}_0  \ket{\chi^{(0)}},
 \end{aligned}
\end{equation}
and where we indicate with a ``bar'' to distinguish from the expansion in terms of the right-eigenvector.
Truncating at second order at most one obtains for the energy
\begin{equation}
 \label{eq:def_e_left_1}
 \begin{aligned}
 &  \bar{E}^{(0)} = \frac{\elemm{\Phi}{\htcz^\dagger}{\chi^{(0)}}}{\braket{\chi^{(0)}}{\Phi}} , \\
 &  \bar{E}^{(1)} = \frac{\elemm{\Phi}{\vtc^\dagger }{\chi^{(0)}}}{\braket{\chi^{(0)}}{\Phi}} , \\
 &  \bar{E}^{(2)} = \frac{\elemm{\Phi}{\vtc^\dagger }{\chi^{(1)}}}{\braket{\chi^{(0)}}{\Phi}} ,
 \end{aligned}
\end{equation}
and the coefficients of the first order perturbed left wave function are
\begin{equation}
 \label{eq:chi_1}
 \begin{aligned}
 \bar{c}_i^{(1)} & = \frac{\elemm{\phi_i}{(\vtc)^\dagger}{\chi^{(0)}}}{\tilde{E}^{(0)} - \epsilon_i} \\
                 & = \frac{\elemm{\chi^{(0)}}{\vtc}{\phi_i}}{\tilde{E}^{(0)} - \epsilon_i}.
 \end{aligned}
\end{equation}

This yields the second order contribution to the left expansion of the energy of the function $\ket{\phi_i}$
\begin{equation}
 \label{eq:def_e_left_2}
 \begin{aligned}
 \bar{E}^{(2)}_i & = \frac{\elemm{\Phi}{(\vtc)^\dagger}{\phi_i}\bar{c}_i^{(1)}}{\braket{\chi^{(0)}}{\Phi}}  \\
                 & =\frac{1}{\braket{\chi^{(0)}}{\Phi}} \frac{\elemm{\chi^{(0)}}{\vtc}{\phi_i} \elemm{\phi_i}{\vtc}{\Phi} }{\tilde{E}^{(0)} - \epsilon_i}.\\
 \end{aligned}
\end{equation}
We see that the in the general case the expansion for the energy in terms of the left- and right-eigenvectors are \textit{a priori} different
as the right-expansion depends on a left-function $\chi$ and the left-expansion depends on a right-function $\Phi$.
Nevertheless, if for the expansion in terms of the right-eigenvector we choose the left-function $\ket{\chi}$
to be the left-eigenvector of $\htcz$ (see Eqs.~\eqref{eq:def_e_right_1}, \eqref{eq:def_e_right_2} and \eqref{eq:def_pt2})
and similarly for the right-function $\ket{\Phi}$ to be $\ket{\Phi^{(0)}}$ in the left-expansion
(see Eqs.~\eqref{eq:def_e_right_1} and \eqref{eq:def_e_right_2}),
one obtains
\begin{equation}
 \begin{aligned}
 &\tilde{E}^{(0)} =\bar{E}^{(0)}  = \frac{\elemm{\chi^{(0)}}{\htcz }{\Phi^{(0)}}}{\braket{\chi^{(0)}}{\Phi^{(0)}}}, \\
 & \tilde{E}^{(1)}= \bar{E}^{(1)} = \frac{\elemm{\chi^{(0)}}{ \vtc}{\Phi^{(0)}}}{\braket{\chi^{(0)}}{\Phi^{(0)}}}, \\
 &\tilde{E}^{(2)}_i=\bar{E}^{(2)}_i = \frac{1}{\braket{\chi^{(0)}}{\Phi^{(0)}}} \frac{\elemm{\chi^{(0)}}{\vtc}{\phi_i} \elemm{\phi_i}{\vtc}{\Phi^{(0)}} }{\tilde{E}^{(0)} - \epsilon_i},
 \end{aligned}
\end{equation}
which means that the right- and left-perturbative expansions for the energy coincide up to second order.
One might nevertheless notice that a function $\ket{\phi_i}$ has potentially different coefficients at first order
in the left- and right-wave function. 

\subsection{Selected CI in a transcorrelated framework}
\label{sec:sci_theo}
After recalling the basics of  Hermitian SCI in Sec.~\ref{sec:hermit_sci},
in this section we explain the main ingredients for the non Hermitian SCI.
In Sec.~\ref{sec:nh_sci} we define the different selection criteria
based on the previous expressions for a perturbative expansion of the non Hermitian problem.
Then in Sec.~\ref{sec:diag_nh}, we explain an iterative scheme to obtain the right- and potentially left-eigenvectors
of a given non Hermitian TC Hamiltonian based only on the usual Hermitian eigensolvers.

\subsubsection{Hermitian SCI in a nutshell}
\label{sec:hermit_sci}
We consider here an iterative SCI algorithm for an Hemitian Hamiltonian, 
in which an iteration $n$ can be summarized as follows.
\begin{enumerate}
 \item A zeroth order set of Slater determinants $\mathcal{P}^n=\{\ket{\rm I},\,i=1,N_{\text{det}} \}$ is known
 and one obtains the ground state eigenvector of the Hamiltonian within $\mathcal{P}^n$
  \begin{equation}
   \begin{aligned}
 &   \ket{\Psi^{(0)}} = \sum_{\rm I\,\in\,\mathcal{P}^n} c_{\rm I}^{(0)} \ket{\rm I} \\
 & E_v \equiv  E^{(0)} = \min_{\{c_{\rm I}^{(0)} \}} \frac{\elemm{\Psi^{(0)}}{\hat{H}}{\Psi^{(0)}}}{\braket{\Psi^{(0)}}{\Psi^{(0)}}}.
   \end{aligned}
  \end{equation}
 As the Hamiltonian is Hermitian, $E^{(0)}$ is necessarily variational and will be referred to as $E_v$ by opposition to $E^{(0)}$ in the case where the Hamiltonian is non Hermitian.
 \item Generate the determinants $\ket{\phi_i} \notin \mathcal{P}^n$ which are connected to $\ket{\Psi^{(0)}}$
  \begin{equation}
   \elemm{\phi_i}{\hat{H}}{\Psi^{(0)}}\ne 0 ,\quad  \phi_i \notin \mathcal{P}^n.
  \end{equation}
 \item For each of these $\ket{\phi_i}$, estimate its \textit{importance criterion} $f_{\phi_i}$.
 \item Select a set of $N_{\phi_i}$ Slater determinants $\ket{\phi_i}$, called $\mathcal{A}_f^n$, with the largest $|f_{\phi_i}|$.
 \item Add the set $\mathcal{A}_f^n$ to $\mathcal{P}^n$ in order to
   define the new set of determinants of the variational space
  \begin{equation}
   \mathcal{P}^{n+1} = \mathcal{P}^n \cup \mathcal{A}_f^n.
  \end{equation}
 \item Go back to step 1) and iterate until a given convergence criterion is reached.
\end{enumerate}
The flavour of SCI essentially changes through the definition of \textit{importance criterion} $f_{\phi_i}$:
it can be for instance the coefficient at first order
\begin{equation}
 f_{\phi_i} \equiv c_{i}^{(1)} = \frac{\elemm{\phi_i}{\hat{H}}{\Psi^{(0)}}}{E^{(0)} -\elemm{\phi_i}{\hat{H}}{\phi_i}},
\end{equation}
the second order contribution to the energy
\begin{equation}
 f_{\phi_i} \equiv E_{i}^{(2)} = \frac{\elemm{\Psi^{(0)}}{\hat{H}}{\phi_i}^2}{E^{(0)} -\elemm{\phi_i}{\hat{H}}{\phi_i}},
\end{equation}
or some minor modifications corresponding to the diagonalization of
the Hamiltonian matrix written in the basis of $\ket{\Psi^{(0)}}$ and
$\ket{\phi_i}$.  
An alternative approach consists in the HCI\cite{hbci} which selects directly the two-electron integrals in the Hamiltonian matrix
in order to screen the generation of the excitation operators which will generate the $\ket{\phi_i}$.

Another important quantity used in SCI is the second order
perturbative contribution to the energy $E^{(2)}$ which can be computed from a set of determinants $\mathcal{P}^n$
\begin{equation}
 E^{(2)} = \sum_{\phi_i \notin \mathcal{P}^n} E_{\phi_i}^{(2)}.
\end{equation}
The FCI ground state energy can be estimated at an iteration $n$ as
\begin{equation}
 E_{\text{FCI}} \approx E_v + E^{(2)}.
\end{equation}
In order to measure the efficiency of a given SCI, one can look at the
rate of convergence of the zeroth order energy $E_v$ and the second order corrected energy $E_v + E^{(2)}$ with respect to
the size of the zeroth order space.

\subsubsection{Non Hermitian SCI: different flavours}
\label{sec:nh_sci}
As the right-eigenvector is supposed to converge faster than the left-eigenvector because of the correlation factor,
it is intuitive to focus on the perturbative expansion of the right-eigenvector $\Phi$.
Nevertheless, as the perturbative expansion of the functional $\etchiphi$ depends
on a general left function, one has to choose which function $\chi$ is used.
One can then distinguish between two flavours of SCI according to if only the right eigenvector is computed
or if one also computes the left-eigenvector.
Therefore, the first change in the non Hermitian SCI consists in the step 1 (the \emph{variational step}): \\
1. Compute the right-eigenvector of $\htc$ within $\mathcal{P}^n$
\begin{equation}
 \begin{aligned}
 & \ket{\Phi^{(0)}}= \sum_{\rm I\,\in\,\mathcal{P}^n} \tilde{c}_{\rm I}^{(0)} \ket{\rm I} \\
 & \htc \ket{\Phi^{(0)}} = E^{(0)} \ket{\Phi^{(0)}},\\
 \end{aligned}
\end{equation}
and potentially also the left-eigenvector
\begin{equation}
 \begin{aligned}
 & \ket{\chi^{(0)}}= \sum_{\rm I\,\in\,\mathcal{P}^n} \bar{c}_{\rm I}^{(0)} \ket{\rm I} \\
 & \htc^\dagger  \ket{\chi^{(0)}} = E^{(0)} \ket{\chi^{(0)}}.\\
 \end{aligned}
\end{equation}
If only the right-eigenvector $\ket{\Phi^{(0)}}$ is computed, then $\chi = {\Phi^{(0)}}$,
and if the left-eigenvector is also computed, then $\chi = {\chi^{(0)}}$. 
Notice that because of the non-Hermitian nature, the zeroth-order energy $E^{(0)}$ is not necessarily variational.  

At this stage, one needs to specify the selection criterion
$f_{\phi_i}$ used to select the determinants $\ket{\phi_i}\notin
\mathcal{P}^n$, and there are two main approaches: the $f_{\phi_i}$
which only require
the right-eigenvector and those which also require the left-eigenvector.

Regarding the $f_{\phi_i}$ needing only the right-eigenvector, one can choose the first order coefficient
\begin{equation}
 \label{f_coef}
 \tilde{f}_{\phi_i}^{\text{coef}} \equiv {c}_{i}^{(1)} = \frac{\elemm{\phi_i}{\tilde{H}}{\Phi^{(0)}}}{E^{(0)} -\elemm{\phi_i}{\hat{H}}{\phi_i}},
\end{equation}
or the \textit{symmetric} second order energy where one sets $\ket{\Phi^{(0)}}$ as the left-function in Eq. \eqref{eq:def_pt2}
\begin{equation}
 \label{f_e_sym}
 \tilde{f}_{\phi_i}^{\text{E sym}} \equiv \tilde{E}^{(2)}_{i}
 =\frac{\elemm{\Phi^{(0)}}{\htc}{\phi_i}\elemm{\phi_i}{\htc}{\Phi^{(0)}}}{E^{(0)} -\elemm{\phi_i}{\hat{H}}{\phi_i}}.
\end{equation}
The latter expression for the second order perturbed energy is similar to the Hermitian case
with nevertheless the difference that $\elemm{\Phi^{(0)}}{\htc}{\phi_i}\ne \elemm{\phi_i}{\htc}{\Phi^{(0)}}$.
The selection criteria $\tilde{f}_{\phi_i}^{\text{E sym}}$ in Eq. \eqref{f_e_sym}
is called the \textit{symmetric} second order energy because it depends only on $\ket{\Phi^{(0)}}$.

The zeroth order energy obtained with the selection criteria $\tilde{f}_{\phi_i}^{\text{coef}}$ and $\tilde{f}_{\phi_i}^{\text{E sym}}$
will be referred to as $E^{(0)}_{\text{coef}}$ and $E^{(0)}_{\text{sym}}$, respectively.
Similarly, the corresponding second order corrected energies referred to as
$(E^{(0)}+E^{(2)})_{\text{coef}}$ and $(E^{(0)}+E^{(2)})_{\text{sym}}$
are obtained by adding the symmetrized second order energy $\tilde{E}^{(2)}_{i}$ of Eq. \eqref{f_e_sym}
to the zeroth order energy $E^{(0)}_{\text{coef}}$ and $E^{(0)}_{\text{sym}}$, respectively.

We also introduce two Hermitian selection criteria which are the second order energy with the usual Hamiltonian
and $\ket{\Phi^{(0)}}$ as the zeroth order wave function
\begin{equation}
 \begin{aligned}
 \tilde{f}_{\phi_i}^{\text{regular H}} & =
\frac{\elemm{\Phi^{(0)}}{\hat{H}}{\phi_i}^2 }{\elemm{\Phi^{(0)}}{\hat{H}}{\Phi^{(0)}} - \elemm{\phi_i}{\hat{H}}{\phi_i}},
 \end{aligned}
\end{equation}
and the second order energy based on the symmetrized operator $\frac{1}{2}(\tilde{H}+\tilde{H}^\dagger)$ and $\ket{\Phi^{(0)}}$ as the zeroth order wave function
\begin{equation}
 \begin{aligned}
 \tilde{f}_{\phi_i}^{\text{Hermit}} & =
\frac{\elemm{\Phi^{(0)}}{\frac{1}{2}(\tilde{H}+\tilde{H}^\dagger)}{\phi_i}^2 }
{\elemm{\Phi^{(0)}}{\frac{1}{2}(\tilde{H}+\tilde{H}^\dagger)}{\Phi^{(0)}} - \elemm{\phi_i}{\frac{1}{2}(\tilde{H}+\tilde{H}^\dagger)}{\phi_i}}.
 \end{aligned}
\end{equation}
The zeroth order energy obtained with the $\tilde{f}_{\phi_i}^{\text{regular H}}$
and $\tilde{f}_{\phi_i}^{\text{Hermit}}$ will be referred to as $E^{(0)}_{\text{regular H}}$ and
$E^{(0)}_{\text{Hermit}}$, respectively. We do not compute any second order corrected energy for these two approaches.

If the left-eigenvector $\ket{\chi^{(0)}}$ is computed, one can then compute another selection criterion based on the energy contribution at second order
\begin{equation}
 \label{eq:f_n_sym}
 \begin{aligned}
 \tilde{f}_{\phi_i}^{\text{E n sym}} & = \tilde{E}^{(2)}_{\phi_i} \\
          & = \frac{1}{\braket{\chi^{(0)}}{\Phi^{(0)}}} \frac{\elemm{\chi^{(0)}}{\vtc}{\phi_i} \elemm{\phi_i}{\vtc}{\Phi^{(0)}} }{\tilde{E}^{(0)} - \epsilon_i}.
 \end{aligned}
\end{equation}
The zeroth order energy obtained with the selection criterion $\tilde{f}_{\phi_i}^{\text{E n sym}}$
will be referred to as $E^{(0)}_{\text{n sym}}$, and the corresponding second order corrected energy
is referred to as $(E^{(0)}+E^{(2)})_{\text{n sym}}$.

\subsubsection{Obtaining left- and right-eigenvectors with iterative Hermitian matrix dressing  }
\label{sec:diag_nh}
As pointed in Sec.~\ref{sec:nh_sci}, SCI in a TC framework requires the right-eigenvectors and potentially also
the left-eigenvectors of large non Hermitian matrices.
This can be done using a non Hermitian variant of the usual Davidson method (see Ref. \onlinecite{CarTruFri-JCTC-10}
and references therein), but we adopt here an alternative strategy based on an iterative Hermitian dressing of the usual Hamiltonian matrix.
Such an approach was originally proposed in Ref. \onlinecite{NebSancMalMay-JCP-95} in the context of single reference coupled cluster
and more recently applied in the context of multi-reference coupled cluster\cite{GinDavSceMal-JCP-16} or quantum Monte Carlo\cite{AmaGinSce-JCTC-22}.

Consider a general non Hermitian operator $\tilde{H}$ which can be decomposed as
\begin{equation}
 \label{eq:eigv_0}
  \tilde{H}  = \hat{H} + \tilde{\Lambda},
\end{equation}
where $\hat{H}$ is Hermitian and $\tilde{\Lambda}$ is non Hermitian.
We target a given right-eigenvector $\ket{\Phi}$ fulfilling the non Hermitian eigenvalue equation
\begin{equation}
 \label{eq:eigv_1}
 \tilde{H} \ket{\Phi} = E\ket{\Phi},\quad \tilde{H}^\dagger \ne \tilde{H},
\end{equation}
and we would like to rewrite this eigenvalue equation as an effective Hermitian problem
\begin{equation}
 \label{eq:heff_0}
 \hat{H}_{\text{eff}} \ket{\Phi} = E\ket{\Phi},\quad \hat{H}_{\text{eff}}^\dagger = \hat{H}_{\text{eff}}.
\end{equation}
Let $\ket{\Phi}$ be decomposed on an orthonormal basis $\{\ket{\rmi} \}$
\begin{equation}
 \ket{\Phi} =  \sum_\rmi c_\rmi \ket{\rmi},
\end{equation}
and let $\rmio$ be the basis function with the largest absolute coefficient $|c_{\rmio}|$.
We project Eq. \eqref{eq:eigv_1} on a basis function $\ket{\rmi}$ and obtain
\begin{equation}
 \label{eq:eigv_2}
 c_{\rmio} \elemm{\rmi}{\hat{H}}{\rmio} + \sum_{\rmj\ne \rmio} c_\rmj \elemm{\rmi}{\hat{H}}{\rmj} + \elemm{\rmi}{\tilde{\Lambda}}{\Phi} = E c_\rmi,
\end{equation}
and similarly on $\ket{\rmio}$
\begin{equation}
 \label{eq:eigv_3}
 c_{\rmio} \elemm{\rmi}{\hat{H}}{\rmio} + \sum_{\rmj\ne \rmio} c_\rmj \elemm{\rmio}{\hat{H}}{\rmj} + \elemm{\rmio}{\tilde{\Lambda}}{\Phi} = E c_{\rmio}.
\end{equation}
We can rewrite Eqs. \eqref{eq:eigv_2} and \eqref{eq:eigv_3} as
\begin{equation}
 \label{eq:eigv_4}
 \begin{aligned}
 & c_{\rmio} \big( \elemm{\rmi}{\hat{H}+\hat{\delta}}{\rmio}  + \big)  + \sum_{\rmj\ne \rmio} c_\rmj \elemm{\rmi}{\hat{H}+\hat{\delta}}{\rmj} = E c_\rmi,\\
 & c_{\rmio} \big( \elemm{\rmi}{\hat{H}+\hat{\delta}}{\rmio}  + \big)  + \sum_{\rmj\ne \rmio} c_\rmj \elemm{\rmio}{\hat{H}+\hat{\delta}}{\rmj} = E c_{\rmio},\\
 \end{aligned}
\end{equation}
where we define the dressing operator $\hat{\delta}$ as the following Hermitian operator
\begin{equation}
 \label{eq:eigv_5}
 \elemm{\rmi}{\hat{\delta}}{\rmj}  = \left\{
    \begin{array}{ll}
 \alpha_{\rmio} & \text{if } \rmj = \rmi = \rmio\\
 \alpha_{\rmi} & \text{if } (\rmj = \rmio, \rmi \ne \rmio) \text{ or } (\rmi = \rmio, \rmj \ne \rmio)\\
 0 &\text{otherwise},
    \end{array}
\right.
\end{equation}
with
\begin{equation}
 \begin{aligned}
 \label{eq:eigv_6}
 &\alpha_{\rmio} =
 \frac{\elemm{\rmio}{\tilde{\Lambda}}{\Phi}}{c_{\rmio}}-\sum_{\rmj \ne \rmio}
 \frac{c_\rmj}{c_{\rmio}}\frac{\elemm{\rmj}{\tilde{\Lambda}}{\Phi}}{c_{\rmio}}, \\
 &\alpha_{\rmi} =
 \frac{\elemm{\rmi}{\tilde{\Lambda}}{\Phi}}{c_{\rmio}} .
 \end{aligned}
\end{equation}
With the definitions of Eqs. \eqref{eq:eigv_4}, \eqref{eq:eigv_5} and \eqref{eq:eigv_6}, one can then define
a non linear effective Hermitian operator $\hat{H}_{\text{eff}}$ as
\begin{equation}
 \label{eq:eigv_7}
 \hat{H}_{\text{eff}} = \hat{H} + \hat{\delta},
\end{equation}
which fulfils Eq. \eqref{eq:heff_0}.
The non linearity of $\hat{H}_{\text{eff}}$  comes from the fact such an operator depends on the solution $\ket{\Phi}$ through
the definitions of Eq. \eqref{eq:eigv_6}.
Notice that the non vanishing matrix elements of the Hermitian
operator $\hat{\delta}$ are only on the row and column corresponding
to $\rmio$.

We can then define an iterative scheme to obtain the right eigenvector $\ket{\Phi}$ of the matrix $\tilde{H}$ by the following procedure.
At an iteration $n$, one assumes that an approximation
 \begin{equation}
\ket{\Phi^{(n)}} = \sum_{\rmi} c_{\rmi}^{(n)} \ket{\rmi}
 \end{equation}
to the unknown right eigenvector $\ket{\Phi}$ is known,
with the corresponding energy
\begin{equation}
 E^{(n)} = \frac{\elemm{\Phi^{(n)}}{\tilde{H}}{\Phi^{(n)}}}{ \braket{\Phi^{(n)}}{\Phi^{(n)}}}.
\end{equation}
From that knowledge, one can then
\begin{enumerate}
 \item build the dressing operator $\hat{\delta}^{(n)}$ by using the coefficients $\{c_{\rmi}^{(n)} \}$ in Eqs.\eqref{eq:eigv_5} and \eqref{eq:eigv_6},
 \item build the Hermitian operator $\hat{H}_{\text{eff}}^{(n+1)} = \hat{H} + \hat{\delta}^{(n)}$ and obtain its eigenvectors
\begin{equation}
 \hat{H}_{\text{eff}}^{(n+1)}  \ket{\Psi_i^{(n+1)}} = E_i^{(n+1)} \ket{\Psi_i^{(n+1)}},
\end{equation}
 \item search for the vector $\ket{\Psi_i^{(n+1)}}$ with the largest overlap with $\ket{\Phi^{(n)}}$ and set it as the new guess $\ket{\Phi^{(n+1)}}$,
 \item iterate until reaching a given convergence criterion on the energy for instance.
\end{enumerate}

When the matrix $\tilde{H}$ is large, which is the typical case of selected CI,
one can of course use a modified Davidson procedure to obtain the
eigenvectors of $\hat{H}_{\text{eff}}^{(n)}$ in 2), as detailed in
Ref~\onlinecite{GarSceGinCaffLoo-JCP-18}.
For a given macro iteration $n$ in the procedure described above,
one has to apply $\hat{H}+\hat{\delta}^{(n)}$ instead of the usual Hamiltonian $\hat{H}$.
The interest of such an approach is that the storage of the $\hat{\delta}^{(n)}$ operator consists only in a vector which has the dimension
of the basis on which $\ket{\Phi}$ is decomposed, which therefore only adds a marginal storage and computational cost
with respect to the usual Davidson algorithm.
Also, if one wishes to obtain the left eigenvector, one just has to replace $\tilde{H}$ by $\tilde{H}^\dagger$ in Eq.~\eqref{eq:eigv_6}.

In the specific case where $\tilde{H}$ is the  TC Hamiltonian matrix written on a given basis of Slater determinants,
we realize the splitting of $\hu$ of Eq. \eqref{eq:eigv_0} as follows:
$\hat{H}$ is the usual Hamiltonian and $\hat{\Lambda}$ all the additional terms arising from the similarity transformation with the correlation factor, \textit{i.e.}
\begin{equation}
 \hat{\Lambda} = -\hat{K} -\hat{L}.
 \label{eq:delta_J}
\end{equation}

\section{Results}
\label{sec:results}
\subsection{Computational details}
Thorough this work, all computations were made using canonical HF MOs.
As all electron calculations are carried, we used the core-valence Dunning atomic orbital (AO) basis family cc-pCVXZ~\cite{WooDun-JCP-95}.
The SCI energies, both with the usual and TC Hamiltonian, reported in the various tables were obtained with a zeroth order space large enough to make
the absolute value of the second order perturbation smaller than $0.5\, \text{mH}$.
Regarding the integrals involved in $\hmub$, the scalar two-body part is computed analytically while
the non-Hermitian and three-body parts are computed using a mixed analytical-numerical scheme utilizing
a Becke numerical grid\cite{Bec-JCP-88b} with 30 radial points and a Lebedev angular grid of 50 grid points.
Numerical tests have shown that these relatively small number of grid points ensures a sub $\mu\text{Ha}$
convergence of the total energies.
All along this work, the value of the parameter $\mu$ used in $\hmub$ is the so-called RSC+LDA as proposed
in Ref. \onlinecite{Gin-JCP-21} in order to compare with the near exact results in a given basis set
obtained with the so-called $\mu$-TC-FCIQMC of Ref. \onlinecite{DobCohAlaGin-JCP-22}.
All CIPSI calculations were carried using the Quantum Package\cite{QP2} and the various TC calculations were obtained using
a plugin created for the Quantum Package.
The equilibrium geometries of the CH$_2$ and FH molecules have been taken from Ref.~\onlinecite{YuaGinTouUmr-JCP-20}
while that of H$_2$O have been taken from Ref.~\onlinecite{CafAplGinSce-JCP-16}.
The estimated CBS all-electron results for atoms are taken from Ref. \onlinecite{ChaGwaDavParFro-PRA-93}.
In the case of molecular systems, the estimated CBS were obtained from the two-point $X^{-3}$ extrapolation of Helgaker \textit{et. al.}~\onlinecite{HelKloKocNog-JCP-97} with $X=\{Q,5\}$ of CIPSI energies, except for H$_2$O for which it was taken from
Ref.~\onlinecite{CafAplGinSce-JCP-16}.

\subsection{Convergence of the different variants of SCI}
\label{sec:conv_sci}
As pointed in Sec.~\ref{sec:nh_sci}, there are several flavours of SCI according to the selection criterion $f_{\phi_i}$ and the left-function $\chi$ chosen.
In this section we investigate the convergence of these different approaches on several atomic and molecular systems in order to find an optimal scheme for both the selection criterion and the left-function. The 5-idx approximation was used for all calculations reported in Sec.~\ref{sec:conv_sci}.

   \begin{figure}
   \centering
\includegraphics[width=0.48\textwidth]{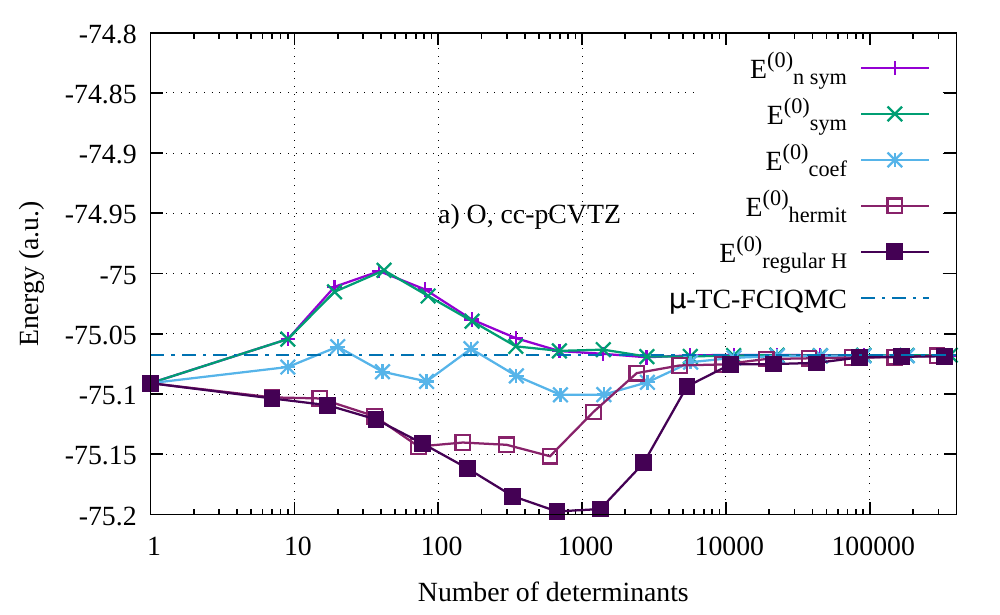}
\includegraphics[width=0.48\textwidth]{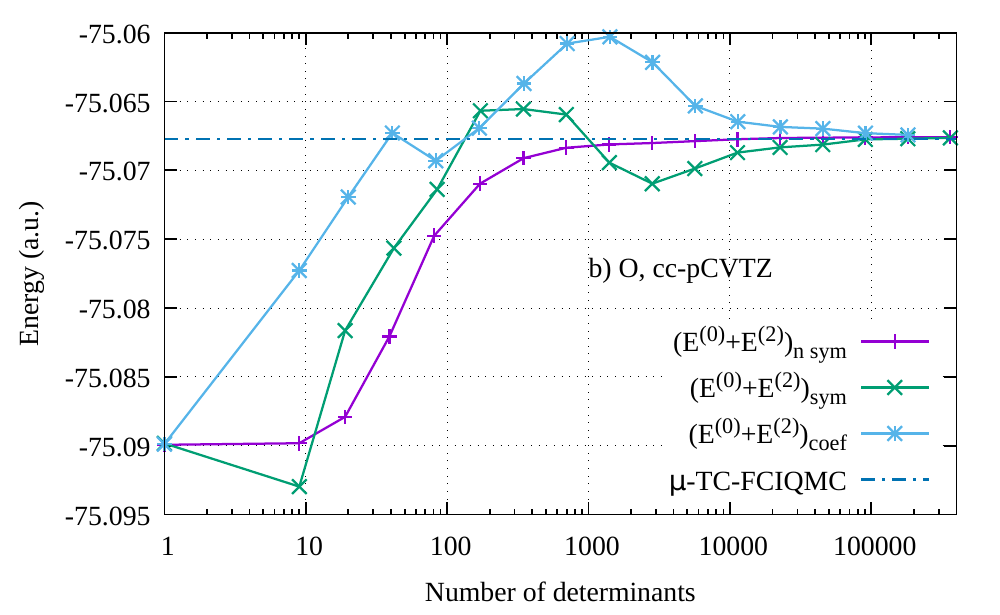}
\includegraphics[width=0.48\textwidth]{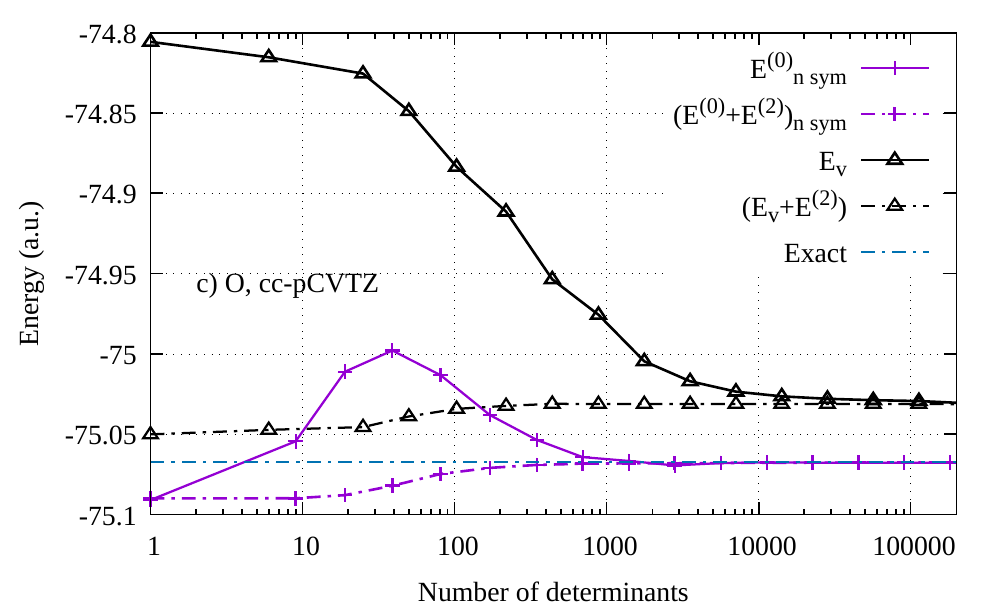}
   \caption{\label{fig_o_tz} O, cc-pCVTZ. Convergence of the zeroth order energy (a), of the corrected energy (b) according to different selection criteria. Comparison with the usual Hermitian CIPSI algorithm in the same basis (c). The exact ground state eigenvalue of the $\mu$-TC Hamiltonian is taken from Ref. \onlinecite{DobCohAlaGin-JCP-22} and referred to as $\mu$-TC-FCIQMC.    }
   \end{figure}

   \begin{figure}
   \centering
\includegraphics[width=0.48\textwidth]{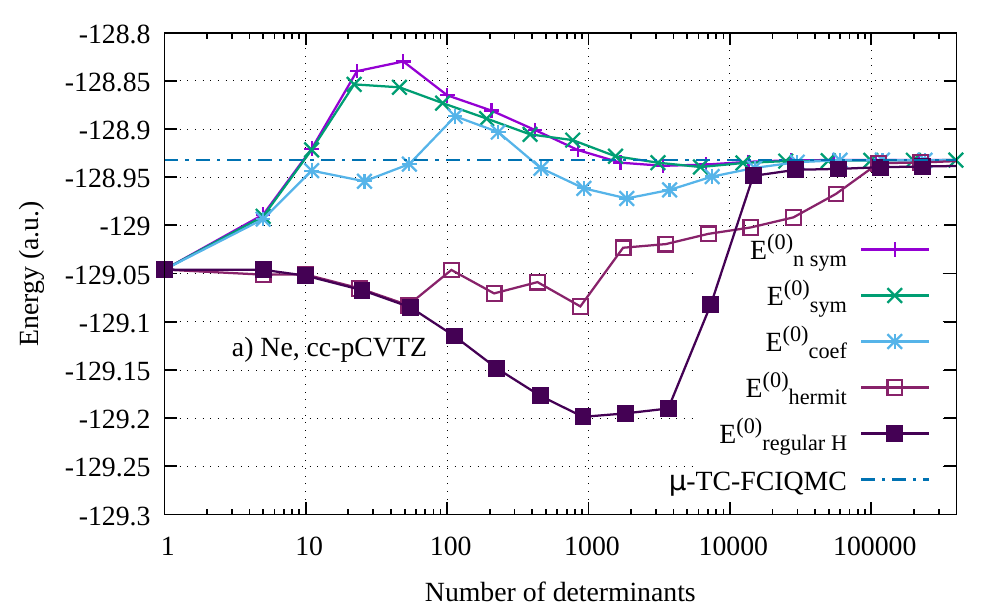}
\includegraphics[width=0.48\textwidth]{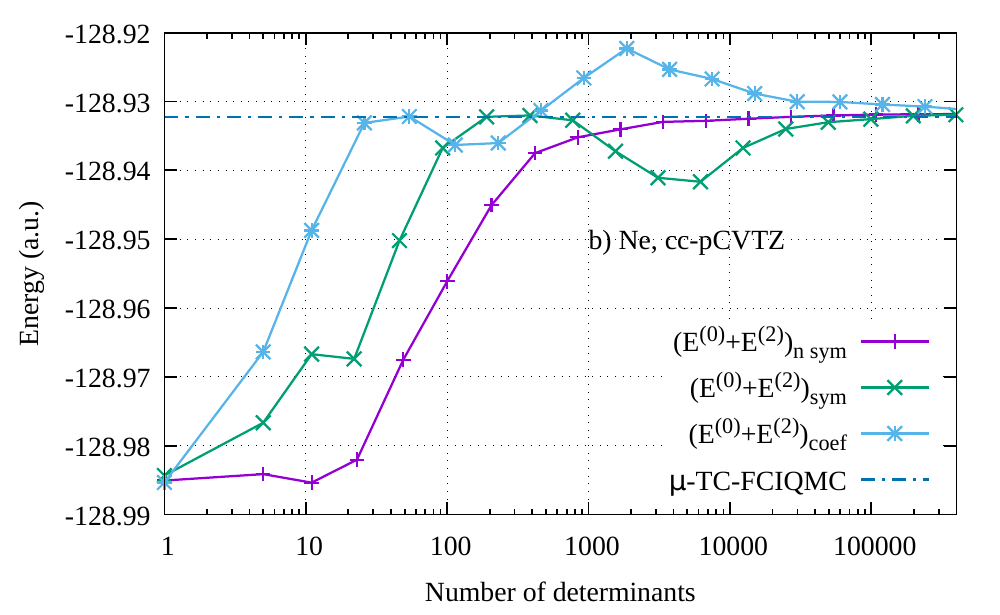}
\includegraphics[width=0.48\textwidth]{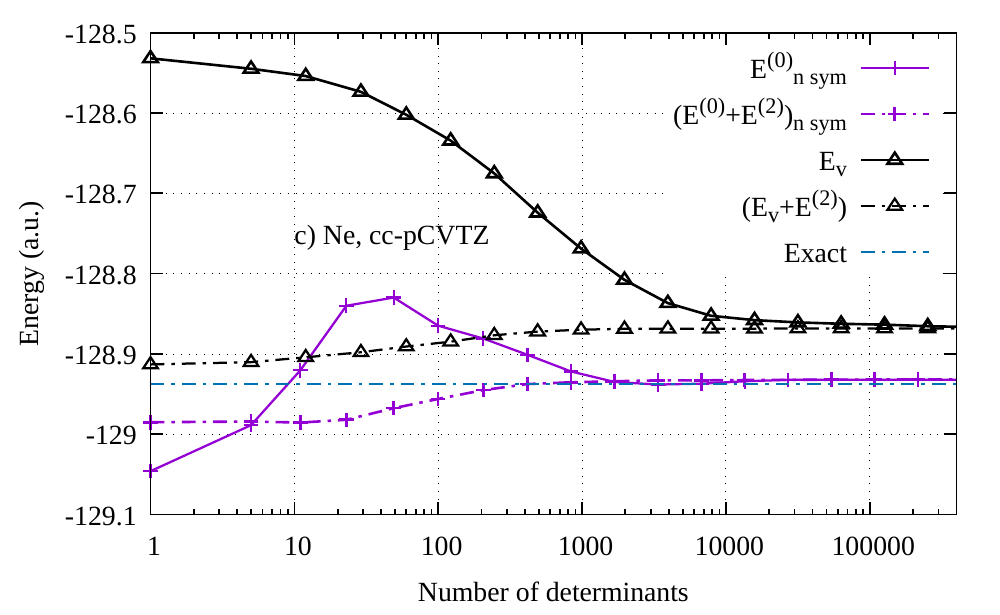}
   \caption{\label{fig_ne_tz} Ne, cc-pCVTZ. Convergence of the zeroth order energy (a), of the corrected energy (b) according to different selection criteria. Comparison with the usual Hermitian CIPSI algorithm in the same basis (c). The exact ground state eigenvalue of the $\mu$-TC Hamiltonian is taken from Ref. \onlinecite{DobCohAlaGin-JCP-22} and referred to as $\mu$-TC-FCIQMC. }
   \end{figure}

   \begin{figure}
   \centering
\includegraphics[width=0.48\textwidth]{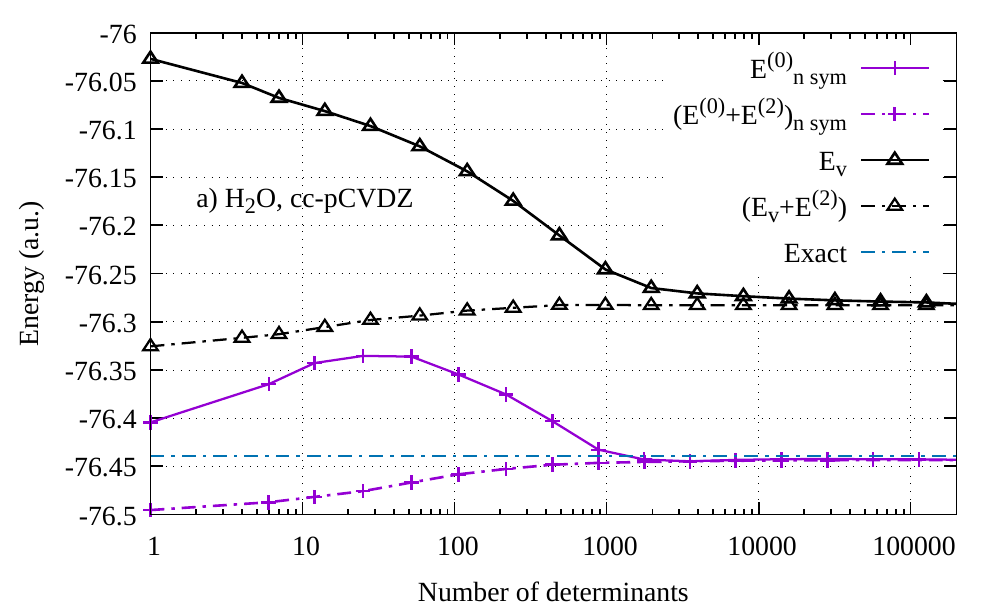}
\includegraphics[width=0.48\textwidth]{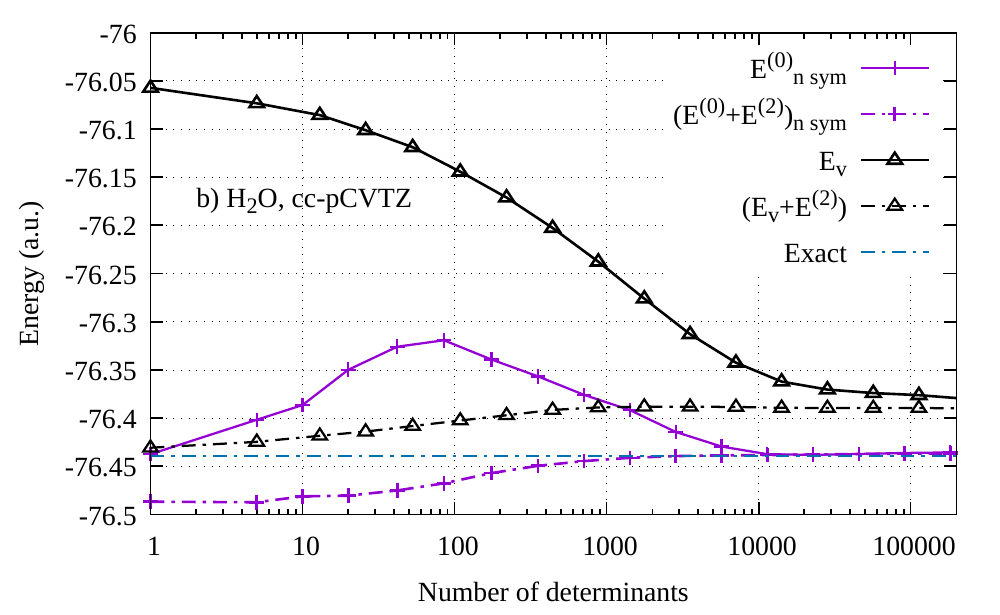}
   \caption{\label{fig_h2o} H$_2$O.  Convergence of the zeroth order energy and the corrected energy
 using the selection criterion $\choicece$ and comparison with the usual Hermitian CIPSI algorithm in the same basis in the cc-pCVDZ (a) and cc-pCVTZ (b) basis sets. The estimated exact energy is taken from the extrapolated DMC energy of Caffarel \textit{et. al.} (see Ref. \onlinecite{CafAplGinSce-JCP-16}).}
   \end{figure}

We report in Fig. \ref{fig_o_tz}, \ref{fig_ne_tz} and \ref{fig_h2o} the convergence of the zeroth order energy together
with the second order corrected energy for the different selection criteria.
Comparison with the usual CIPSI algorithm in the same basis are also reported.
Focusing first on the zeroth order energy, we can observe that
i) the non variational character of the TC SCI is manifest as the zeroth order energy can be way below the exact value within a basis set which is estimated by the corresponding $\mu$-TC-FCIQMC energy in the basis set,
ii) the convergence is non monotonic for all selection criteria, which is also a sign of the non Hermitian character of the TC Hamiltonian,
iii) the zeroth order energies obtained with the two selection criteria using only Hermitian quantities
(\textit{i.e.} $\choiceh$ and $\choicehht$) converge much slower than those obtained with
the selection criteria using non Hermitian quantities (\textit{i.e.} $\choiceae$, $\choicebe$ and $\choicece$),
iv) the zeroth order energies obtained with the criteria based on the first order coefficient (\textit{i.e.}
$\choiceae$) converge significantly less rapidly than
those based with second order energy contribution (\textit{i.e.} $\choicebe$ and $\choicece$),
v) the zeroth order energies obtained with the criteria $\choicebe$ and $\choicece$ converge essentially in the same way,
vi) the zeroth order energy tends to converge faster than the usual variational energy in the Hermitian CIPSI algorithm although usual HF orbitals are used which would favor the usual Hermitian calculations.

Turning now on the convergence of the second order corrected energy, we can observe that
i) the convergence of $\choiceaept$ and $\choicebept$ are non monotonic,
ii) the convergence using the criterion $\choicecept$ is much faster and stable than that of criterion $\choiceaept$ and $\choicebept$,
iii) the convergence of $\choicecept$ is comparable to that of the usual second order corrected energy in the Hermitian CIPSI algorithm,
iv) when reaching convergence, the zeroth order and second order corrected energies agree very well with the estimated exact $\mu$-TC-FCIQMC of Ref.~\onlinecite{DobCohAlaGin-JCP-22}, which is expected as one reaches the TC-FCI limit.
We can notice that the point ii) illustrates the importance of having a left function as close as possible
from the exact left-eigenvector of the TC Hamiltonian in the basis.

Based on these results we can conclude that
taking into account the non Hermitian character of $\hu$ is mandatory to actually select the important determinants in the context of the TC Hamiltonian. 
Selecting on the energy leads to a significant improvement of the convergence of the zeroth order energy with respect to the selection based on the first order coefficient
(which is not necessarily the case in the Hermitian case),
and having a left function as close as possible to the left-eigenvector strongly improves the convergence and stability
of the second order corrected energy. Therefore, $\tilde{f}_{\phi_i}^{\text{E n sym}}$ of Eq. \eqref{eq:f_n_sym}
is the best choice for the selection criterion.
From thereon, all SCI in the TC context are done using the $\tilde{f}_{\phi_i}^{\text{E n sym}}$ selection criterion and
the $\ket{\chi^{(0)}}$ as the left-function $\chi$, which will be referred to as $\mu$-TC-CIPSI.

\subsection{Dependence of total energies and energy differences with the treatment of three-body terms}
\label{sec:res_three_body}

\begin{table}[t]
\caption{\label{3_body} Total energies (a.u.) for several atomic and molecular systems at different levels of approximations.
}
\begin{tabular}{clll}
\toprule
       &  cc-pCVDZ & cc-pCVTZ & cc-pCVQZ \\
\hline
\multicolumn{4}{c}{Carbon} \\
CIPSI                                    &-37.79798  &-37.83003    &-37.83962          \\
\hline
$\mu$-TC-FCIQMC                          & -37.84888 & -37.84793   &-37.84617          \\
$\mu$-TC-CIPSI(5-idx)                    & -37.84875 & -37.84809   &                   \\
$\mu$-TC-CIPSI(4-idx)                    & -37.84901 & -37.84837   &-37.84630          \\
\multicolumn{4}{c}{Estimated exact} \\
\multicolumn{4}{c}{-37.84500}\\
\hline
\hline
\multicolumn{4}{c}{Oxygen} \\
CIPSI                                    & -74.95051 &  -75.03122  &  -75.05447        \\
\hline
$\mu$-TC-FCIQMC                          & -75.07412 &  -75.06774  &  -75.06729        \\
$\mu$-TC-CIPSI(5-idx)                    & -75.07420 &  -75.06759  &  -75.06750        \\
$\mu$-TC-CIPSI(4-idx)                    & -75.07420 &  -75.06761  &  -75.06744        \\
\multicolumn{4}{c}{Estimated exact} \\
\multicolumn{4}{c}{-75.06730}\\
\hline
\hline
\multicolumn{4}{c}{Fluorine} \\
CIPSI                                    & -99.56965 & -99.68185   & -99.71509         \\
\hline
$\mu$-TC-FCIQMC                          & -99.74701 & -99.73164   & -99.73326         \\
$\mu$-TC-CIPSI(5-idx)                    & -99.74695 & -99.73105   & -99.73282         \\
$\mu$-TC-CIPSI(4-idx)                    & -99.74725 & -99.73172   & -99.73345         \\
\multicolumn{4}{c}{Estimated exact} \\
\multicolumn{4}{c}{-99.73390}\\
\hline
\hline
\multicolumn{4}{c}{Neon} \\
CIPSI                                    & -128.72254           & -128.86823  &   -128.91235 \\
\hline
$\mu$-TC-FCIQMC                          & -128.96435           & -128.93221  &   -128.93569 \\
$\mu$-TC-CIPSI(5-idx)                    & -128.96437           & -128.93180  &   -128.93570 \\
$\mu$-TC-CIPSI(4-idx)                    & -128.96496           & -128.93321  &   -128.93712  \\
\multicolumn{4}{c}{Estimated exact} \\
\multicolumn{4}{c}{-128.93760}\\
\hline
\hline
\multicolumn{4}{c}{H$_2$O} \\
CIPSI                                    & -76.28287\phantom{(20)}    &-76.38993\phantom{(20)}    & -76.42156        \\
\hline
$\mu$-TC-FCIQMC                          & -76.44338            &-76.43700   &  -                             \\
$\mu$-TC-CIPSI(5-idx)                    & -76.44354            &-76.43713   &  -               \\
$\mu$-TC-CIPSI(4-idx)                    & -76.44399            &-76.43750   & -76.43983        \\
\hline
\multicolumn{4}{c}{Estimated exact} \\
\multicolumn{4}{c}{-76.43894(12)}\\
\hline
\hline
\multicolumn{4}{c}{CH$_2$} \\
CIPSI                                    & -39.06130                  &   -39.11232               & -39.12689        \\
\hline
$\mu$-TC-FCIQMC                          & -39.13076                  &   -39.13614               &  -               \\
$\mu$-TC-CIPSI(5-idx)                    & -39.13074                  &   -39.13658               &  -               \\
$\mu$-TC-CIPSI(4-idx)                    & -39.13105                  &   -39.13685               & -39.13520        \\
\hline
\multicolumn{4}{c}{Estimated exact} \\
\multicolumn{4}{c}{-39.13425    }\\
\hline
\hline
\multicolumn{4}{c}{FH} \\
CIPSI                                    & -100.27094                 &   -100.40010              & -100.43818       \\
\hline
$\mu$-TC-FCIQMC                          & -100.47331                 &   -100.45555              &  -               \\
$\mu$-TC-CIPSI(5-idx)                    & -100.47334                 &   -100.45528              &  -               \\
$\mu$-TC-CIPSI(4-idx)                    & -100.47394                 &   -100.45651              & -100.45959       \\
\hline
\multicolumn{4}{c}{Estimated exact} \\
\multicolumn{4}{c}{-100.46008   }\\
\botrule
\end{tabular}
\end{table}

\begin{table}
\caption{\label{mol_table_deltae}Atomization energy (a.u.) with different methods.
}
\begin{tabular}{clll}
\toprule
                                         & cc-pCVDZ &  cc-pCVTZ  & cc-pCVQZ      \\
\hline
\multicolumn{4}{c}{H$_2$O} \\
CIPSI                                    &   333.80 &   359.09   &  367.19 \\
\hline
$\mu$-TC-FCIQMC                          &   370.7  &   369.64   &     -    \\
$\mu$-TC-CIPSI(5-idx)                    &   370.62 &   369.92   &     -   \\
$\mu$-TC-CIPSI(4-idx)                    &   371.23 &   370.00   &  372.50  \\
\multicolumn{4}{c}{Estimated exact} \\
\multicolumn{4}{c}{371.64(12)}\\
\hline
\hline
\multicolumn{4}{c}{CH$_2$} \\
CIPSI                                    & 264.77   & 282.67     & 287.38   \\
\hline
$\mu$-TC-FCIQMC                          & 283.32   & 288.41     &   -      \\
$\mu$-TC-CIPSI(5-idx)                    & 283.43   & 288.87     &   -      \\
$\mu$-TC-CIPSI(4-idx)                    & 283.48   & 288.87     & 289.01   \\
\multicolumn{4}{c}{Estimated exact} \\
\multicolumn{4}{c}{289.22     }\\
\hline
\hline
\multicolumn{4}{c}{FH    } \\
CIPSI                                    &  202.0   & 218.48     & 223.15   \\
\hline
$\mu$-TC-FCIQMC                          &  227.02  & 224.10     &   -      \\
$\mu$-TC-CIPSI(5-idx)                    &  227.11  & 224.42     &   -      \\
$\mu$-TC-CIPSI(4-idx)                    &  227.41  & 224.98     & 226.16   \\
\multicolumn{4}{c}{Estimated exact} \\
\multicolumn{4}{c}{226.18     }\\
\botrule
\end{tabular}
\end{table}

Having established the best algorithm for non Hermitian SCI, we now focus on the quality of the
computed energies and how they depend on the treatment of the three-body terms.
We study second-row atomic and molecular systems as in Ref.~\onlinecite{DobCohAlaGin-JCP-22}.

We report in Table~\ref{3_body} the total energies obtained for several atomic and molecular systems
using our $\mu$-TC-CIPSI scheme within the 5-idx and 4-idx treatment of the three body terms and
we compare to the $\mu$-TC-FCIQMC results of Ref.~\onlinecite{DobCohAlaGin-JCP-22} which correspond
to near exact results within a basis set with full treatment of the three-body terms.
Computation of atomization energies for molecular systems are reported in Table~\ref{mol_table_deltae}.

From Table~\ref{3_body} and  Table~\ref{mol_table_deltae} one can notice that
both the total energies and atomization energies obtained using the $\mu$-TC-CIPSI and $\mu$-TC-FCIQMC are much closer
to the exact non relativistic energy than the usual CIPSI energies in the same basis set
which illustrates the benefit of having a correlation factor.
The absolute energy difference between the total energies with the full treatment of the three body ($\mu$-TC-FCIQMC)
and that with the 5-idx approximation is never larger than 1~mH for the atomic and molecular systems studied here,
and the absolute difference between the total energies with the $\mu$-TC-FCIQMC and that with the 4-idx approximation is slightly larger
(the largest difference being of 1.5~mH in the case of the Neon atom in the cc-pCVQZ basis set). This difference 
tends to increase with the nuclear charge.
The 4-idx approximation tends to lead to lower energies than the $\mu$-TC-FCIQMC or 5-idx energies,
and the absolute difference between the atomization energies computed with the $\mu$-TC-FCIQMC and that computed with the 5-idx or 4-idx approximation is never larger that 0.5~mH for all molecular systems studied here.

Based on these results we can conclude that the 5-idx and 4-idx approximations weakly affect here the results,
which is encouraging considering the considerable saving in memory storage with respect to the full treatment of the three-body terms.

\section{Conclusion}
\label{sec:conclu}
The present work is dedicated to the extension of the popular SCI algorithm to the TC Hamiltonian.
The main focus of this work is not to study the quality of the results with respect to the specific
correlation factor used but rather to investigate what are the new features of SCI in the TC framework.
We therefore use a rather simple one-parameter correlation factor\cite{Gin-JCP-21,DobCohAlaGin-JCP-22} (see Sec.~\ref{sec:mu_tc}) which
reproduces most of the features of the TC framework, such as a faster convergence convergence
towards CBS results and the non Hermitian property.
The connection between non-Hermitian eigenvalue problems and the search of stationary points of functionals depending on
general left- and right-functions (see Sec.~\ref{sec:n_h_theo}) allows us to propose different choices for both the selection criterion and the second order perturbation energy,
which are central ingredients in the context of SCI.
Based on the numerical investigations performed here (see Sec.~\ref{sec:conv_sci}), we found that i) the selection of the important Slater determinants is strongly affected by the presence of the correlation factor,
ii) taking into account the non Hermitian character of the TC Hamiltonian is mandatory to obtain a fast convergence of the TC energy,
iii) not like in usual SCI, selection criteria based on the first order coefficient or second order energy lead
to significantly different convergence rates of the TC energy,
iv) within a given determinant space, the use of both the left- and right-eigenvectors is mandatory to obtain a smooth
convergence of the second order perturbed energy.
The variational step in SCI transforms into the TC framework in obtaining both left- and right eigenvectors of
large non symmetric matrices, which requires additional non-Hermitian eigensolver technology.
In the present work the latter aspect is by-passed using only Hermitian algorithms thanks to
the use of a low-memory footprint self-consistent dressing\cite{NebSancMalMay-JCP-95,GinDavSceMal-JCP-16, AmaGinSce-JCTC-22} of the usual Hamiltonian.
Within the near-optimal set-up proposed here, we found that the TC-SCI expansion converges faster both in terms of number of Slater determinants and basis set size.
We also investigated the dependence of the results with the level of treatment of the three-body terms and introduced a new approximation,
the 4-idx, which has a typical $N^4$ scaling.
The numerical results obtained here show that this approximation weakly affect the quality of the results both for total energies and energy differences.
We believe that this work opens the way to obtain even more efficient SCI algorithms
with smaller basis set truncation errors.
The role of the orbitals used for the SCI expansion was nevertheless not investigated here, although it is clear that using better adapted MOs will play an important
role in improving the convergence of the energy. On-going work will address this aspect in details,
especially regarding the use of bi-orthonormal basis sets.

\begin{acknowledgments}
This work was performed using HPC resources from GENCI-TGCC
(gen1738,gen12363) and from CALMIP (Toulouse) under allocation
P22001, and was also supported by the European Centre of
Excellence in Exascale Computing TREX --- Targeting Real Chemical
Accuracy at the Exascale. This project has received funding from the
European Union's Horizon 2020 --- Research and Innovation program ---
under grant agreement no.~952165.
\end{acknowledgments}


%
 \end{document}